\documentclass[12pt,draftclsnofoot,onecolumn]{IEEEtran}

\usepackage[utf8]{inputenc}
\usepackage{cite}
\usepackage{floatrow}
\usepackage{amsmath,amssymb,amsfonts,stfloats}
\usepackage{tabularx}
\usepackage{algorithmic}
\usepackage{graphicx}
\usepackage{textcomp}
\usepackage{floatrow}
\usepackage{xcolor}
\usepackage{booktabs}

\usepackage{multicol}
\usepackage{acronym}
\usepackage{url}
\usepackage{mathtools}
\usepackage[font=small]{caption}
\usepackage[caption=false,font=footnotesize]{subfig}

\def\BibTeX{{\rm B\kern-.05em{\sc i\kern-.025em b}\kern-.08em
    T\kern-.1667em\lower.7ex\hbox{E}\kern-.125emX}}

\begin{document}

\title{Reconfigurable and Static EM Skins \\ on Vehicles for Localization
\thanks{This work was partially supported by the European Union under the Italian National Recovery and Resilience Plan (NRRP) of NextGenerationEU, partnership on “Telecommunications of the Future” (PE00000001 - program “RESTART”).} 
}

\author{Dario~Tagliaferri,~\IEEEmembership{Member,~IEEE}, Marouan~Mizmizi,~\IEEEmembership{Member,~IEEE}, Giacomo~Oliveri,~\IEEEmembership{Senior Member,~IEEE}, Umberto~Spagnolini,~\IEEEmembership{Senior Member,~IEEE} and Andrea~Massa,~\IEEEmembership{Fellow,~IEEE} 
\thanks{D.\, Tagliaferri, M.\, Mizmizi, U.\, Spagnolini are with the  Department of Electronics, Information and Bioengineering (DEIB) of Politecnico di Milano, 20133 Milan, Italy  (e-mail: [dario.tagliaferri, marouan.mizmizi, umberto.spagnolini]@polimi.it }
\thanks{U.\, Spagnolini is Huawei Industry Chair}
\thanks{G. Oliveri is with the ELEDIA Research Center (ELEDIA@UniTN - University of Trento), DICAM - Department of Civil, Environmental, and Mechanical Engineering, Via Mesiano 77, 38123 Trento - Italy (email: giacomo.oliveri@unitn.it)}
\thanks{A. Massa is with the ELEDIA Research Center (ELEDIA@UESTC-UESTC), School of Electronic Science and Engineering, University of Electronic Science and Technology of China (UESTC), Chengdu 611731, China, also with the ELEDIA Research Center (ELEDIA@UniTN - University of Trento), DICAM - Department of Civil, Environmental, and Mechanical Engineering, Via Mesiano 77, 38123 Trento - Italy, also with the ELEDIA Research Center (ELEDIA@TSINGHUA - Tsinghua University), Beijing100084, China, also with the School of Electrical Engineering, Tel Aviv University, Tel Aviv 69978, Israel  (e-mail: andrea.massa@unitn.it).}}

\maketitle

\begin{abstract}
Electromagnetic skins (EMSs) have been recently considered as a booster for wireless sensing, but their usage on mobile targets is relatively novel and could be of interest when the target reflectivity can/must be increased to improve its detection or the estimation of parameters. In particular, when illuminated by a wide-bandwidth signal (e.g., from a radar operating at millimeter waves), vehicles behave like \textit{extended targets}, since multiple parts of the vehicle's body effectively contribute to the back-scattering. Moreover, in some cases perspective deformations challenge the correct localization of the vehicle. To address these issues, we propose lodging EMSs on vehicles' roof to act as high-reflectivity planar retro-reflectors toward the sensing terminal. The advantage is twofold: \textit{(i)} by introducing a compact high-reflectivity structure on the target, we make vehicles behave like \textit{point targets}, avoiding perspective deformations and related ranging biases and \textit{(ii)} we increase the reflectivity the vehicle, improving localization performance. We detail the EMS design from the system-level to the full-wave-level considering both reconfigurable intelligent surfaces (RIS) and cost-effective static passive electromagnetic skins (SP-EMSs). Localization performance of the EMS-aided sensing system is also assessed by Cramér-Rao bound analysis in both narrowband and spatially wideband operating conditions. 

\end{abstract}



\section{Introduction}\label{sect:intro}

Localization is a fundamental pre-requisite for many applications and use-cases of the sixth generation of communication systems (6G), that is expected to enable novel verticals such as automated driving, augmented reality, high-fidelity digital twins of the physical environment and others \cite{9349624}. High-accuracy localization (i.e., below $10$ cm of accuracy \cite{9815783}) is supported by the release of large bandwidths in the millimeter wave (mmWave) and sub-THz portions of the EM spectrum ($< 300$ GHz) \cite{9625032}, as well as by the use of massive antenna arrays, either dedicated to sensing purposes (e.g., radars) or already in place for communication.   
Electromagnetic skins (EMSs)  are surging as one of the killer technologies for 6G, boosting communication performance as well as localization one, as recently shown \cite{di2020smart}. In this framework, several different technologies are emerging to implement this functionality, including reconfigurable intelligent surfaces (RISs). Such devices consists of collection of sub-wavelength-sized passive elements, called \textit{meta-atoms}, whose complex reflection coefficients can be electronically tuned to manipulate the incident and reflected/refracted wavefronts \cite{direnzo2021surveyRIS}. From the localization perspective, RISs can be designed to operate as anomalous mirrors to extend the range in non-line-of-sight (NLOS) \cite{10044963} or to improve the angular diversity by magnifying the transmitter (Tx) or receiver (Rx) aperture \cite{8264743}, boosting the estimation of position and orientation of one or multiple targets. In the following, we revise the state of the art on EMSs used for localization and we then outline the contributions of this work.


\begin{figure}[!b]
    \centering
    \includegraphics[width=0.5\columnwidth]{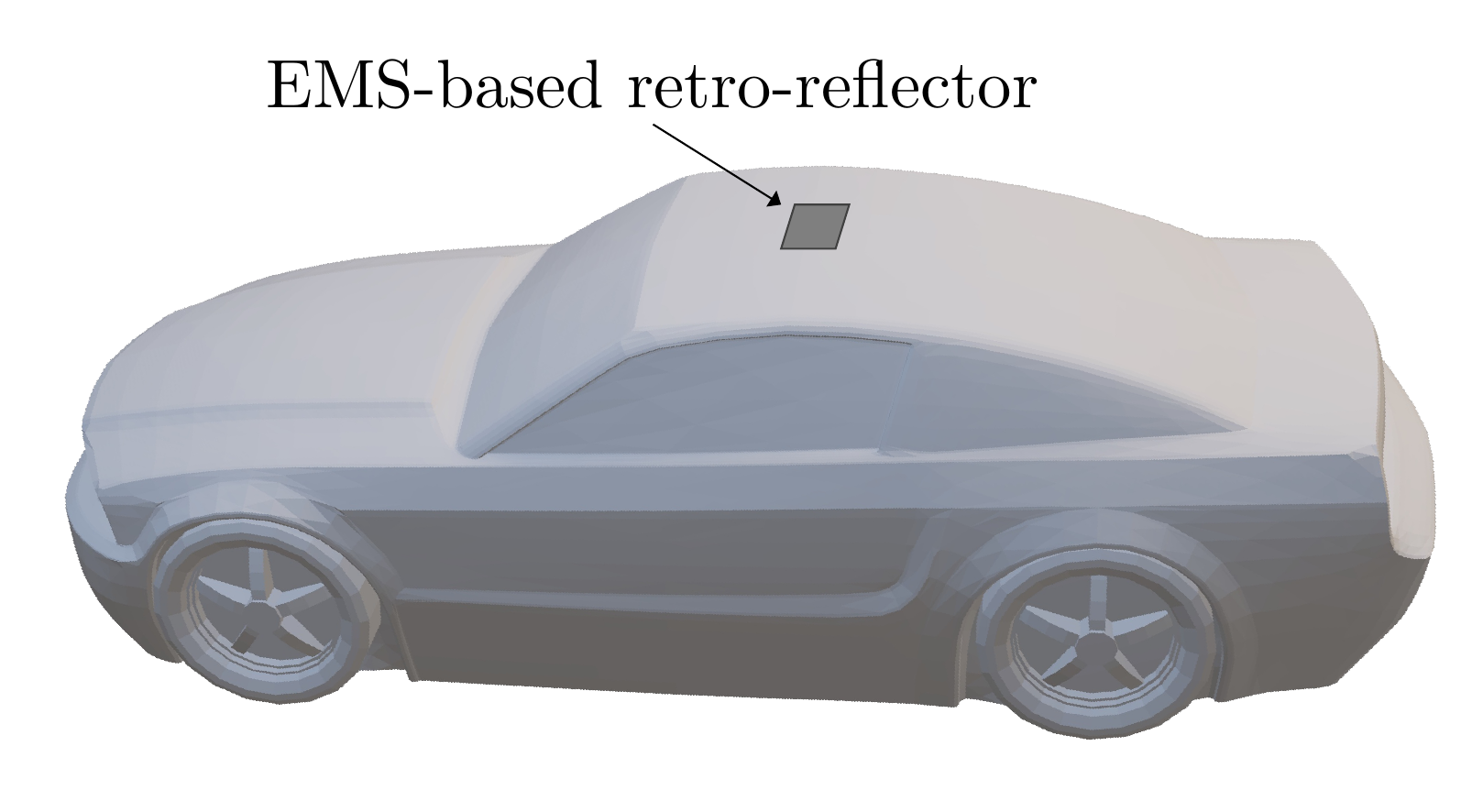}
    \caption{EMS-based retro-reflector on the roof of the vehicle for localization purposes}
    \label{fig:idea}
\end{figure}

\subsection{Literature survey}

The literature on RIS-aided localization underwent a significant growth in the last few years \cite{9775078}. In~\cite{Buzzi_RISforradar_letter,Buzzi_RISforradar_journal}, the authors propose to use a RIS to assist radar in NLOS conditions, addressing on the maximization of the probability of correct detection of a given target, via signal-to-noise (SNR) maximization with constrained phase design. The results analyze the Cramér Rao bound (CRB) on position estimation and suggest the placement of the RIS nearby either the radar or the target, to combat the severe path loss induced by a double or triple reflection (Tx-RIS-target-RIS-Rx). The same authors extend the previous works to the usage of active RIS, hence taking into account architectures with amplification capabilities \cite{Buzzi2022_activeRIS}. The authors of \cite{9508872} analyze the CRB-derived position and orientation error bounds on a RIS-aided localization system, showing numerical results and proposing a suitable CRB-minimizing phase design approach. Radar surveillance in NLOS scenarios assisted by RIS is proposed in \cite{9511765}, whereby different operating regimes of the RIS are discussed. The works \cite{Zhang2022metalocalization,Zhang2022metaradar} consider indoor user localization by means of RIS and received signal strength measurements. Paper \cite{Wang2022_location_awareness} compares the localization performance of RIS-aided and continuous EMSs-aided integrated sensing and communication (ISAC) systems, deriving the CRB in both cases. The paper \cite{Dardari2023_twotimescale} proposes to adopt a two-timescale phase design approach in ISAC systems to reduce the signaling overhead to acquire the channel state information (CSI) at the RIS. RIS phases are kept constant during the location coherence interval of the user/target while the precoder at the base station is fast varying according to the communication channel coherence time. The authors of \cite{10096904} address the problem of RIS modeling mismatch in RIS-aided localization, namely deriving the misspecified CRB on positioning performance for the case in which the system assumes the RIS to have a phase-only control (and unit amplitude) while the true reflection coefficient has non-negligible amplitude variation with the incidence angle. 
Recently, research advances on EMSs shifted towards large implementations (whose size is comparable with the propagation distance) to boost both communication and localization performance by increasing the effective aperture \cite{9838638}. With the increased aperture, literature works started to explore the challenges and possibilities of the near-field operation, namely the additional degrees of freedom brought by a non-planar wavefront across the RIS \cite{10146296}.  The authors of \cite{9625826} derive the fundamental position error bounds for near-field RIS-aided localization systems. The paper \cite{9650561} addresses the near-field target localization in NLOS scenarios, evaluating the CRB on position estimation and proposing an optimization problem to design the RIS phases for CRB minimization. The work \cite{9709801} exploits the near-field-operating RIS as a localization enabler for a single-antenna terminal that is willing to localize a target in the environment. The authors deal with the multipath problem by leveraging on compressed sensing algorithms. The authors of \cite{9921216} consider an hybrid RIS, where a limited number of RF chains at the RIS enables basic signal processing on the received signal.

Previous works assume that RISs are statically placed in the environment, e.g., on the building facades, magnifying the Tx/Rx aperture to improve the localization accuracy. Very few works consider the possibility of directly placing the RIS on targets, to facilitate their localization \cite{Wymeersch2021,Wymeersch2022,Zhang2023_targetRIS}. For instance, the authors of \cite{Wymeersch2021} exploit one Tx terminal and multiple asynchronous Rx stations to localize users equipped with RIS via time-of-arrival estimation. A phase design is proposed and the simulation results are compared with the CRB on 3D position estimation. The same authors extended the previous work to near-field operating conditions in \cite{Wymeersch2022}. More recently, the authors of \cite{Zhang2023_targetRIS} consider the estimation of position and orientation of a target-mounted RIS (lodged on an unmanned aerial vehicle), deriving the CRB on position and orientation. The aforementioned works only considered a \textit{narrowband} RIS operation, namely a frequency-flat RIS reflection model over the whole Tx signal bandwidth. In most sensing scenarios, the employed bandwidth by the active Tx terminal is large enough to induce \textit{wideband} effects at the RIS, namely a frequency-selective RIS response within the bandwidth of interest. Wideband effects give rise to reflection beam squinting/defocusing (depending on the far-/near-field operating condition) as shown in  \cite{Wymeersch2022_wideband_RIS,tagliaferri2023wideband}, therefore current literature on target-lodged RIS does not fully capture the RIS reflection behavior for very large bandwidth localization systems.

\subsection{Contribution}

This paper focuses on the usage of EMSs to improve vehicles' localization when illuminated from above by a large-bandwidth (GHz-wide) sensing signal, such as ISAC terminal or a multiple-input-multiple-output (MIMO) radar. Lodging an EMS on a vehicle has been explored in our previous works \cite{Mizmizi2022_conformal,Tagliaferri2022_conformal6gnet,Mizmizi2022_conformal_GLOBECOM}, for communication purposes only. When the resolution cell of the sensing system is much smaller than the vehicle size, the latter behaves like an \textit{extended target}, challenging its localization and tracking \cite{Wymeersch_CRB_radar_extendedvehicle,9399297}. Moreover, when the sensing terminal is above the vehicle, perspective deformations such as overlay and foreshortening hinder the correct localization of the vehicle from sensing data, as double bounces with the vehicle and the ground bias the time-of-arrival measurements \cite{naraghi1983geometric}. In this setting, we propose to lodge an EMS on the top side of ground vehicles (Fig. \ref{fig:idea}) to assist radar localization by acting as an omnidirectional retro-reflector. The advantage is twofold: \textit{(i)} the EMS increases the reflectivity of the vehicle, thus its radar cross section (RCS) and hence its \textit{visibility} in sensing data and \textit{(ii)} with a proper EMS design, the vehicle behaves like a \textit{point target}, allowing the detection, localization and tracking of a single, known region of the vehicle (e.g., the position of the mmWave communication transceiver of the roof).  
Although the realization of EMS-based retro-reflectors at mmWave/THz bands are present in literature \cite{Desai2020TerahertzVA}, in this setting, the EMS is operating in mobility, and its dynamic configuration is investigated herein.

The main contributions of the paper are as follows:
\begin{itemize}
\item We motivate the use of EMSs for vehicle localization with an experimental radar data acquisition campaign. We employ a Texas Instrument high-resolution mmWave radar mounted at $8$ m from ground that illuminates a travelling vehicle. Through experiments, we demonstrate the severe perspective deformations in the radar image of the vehicle, and we demonstrate the benefits of having EMSs acting as \textit{EM markers} on vehicles' roofs. 
\item We propose a RIS-based retro-reflector design for vehicles. RIS phases are set to enable intentional retro-reflection of the impinging sensing signal towards the sensing terminal. With the aforementioned experimental tests, we empirically devise the the minimum radar cross section (RCS) required for the RIS (thus its physical size) to make it detectable in the sensing data. Then, we show that an inaccurate RIS phase design, e.g., only based on coarse vehicle's positioning (such as GPS), leads to a drastic reduction of the average RCS of several dBm$^2$ due to reflection misalignment with the impinging sensing signal. To obviate for the latter issue, we detail a dynamic RIS-alignment sweeping procedure that designs the phase of the RIS elements as function of \textit{(i)} the coarse vehicle's position estimate w.r.t. the sensing terminal, \textit{(ii)} the uncertainty on vehicle's position estimate and \textit{(iii)} the beamwidth of the RIS reflection pattern. Mobility considerations are added and discussed.
\item To decrease the implementation cost and complexity of a RIS-based reflector, we propose a novel design based on static-passive EM skins (SP-EMSs), an inexpensive solution (two/three orders of magnitude less w.r.t. RISs) to enable advanced wave manipulation and anomalous reflection capabilities, obtained by tailoring the micro-scale geometric/physical properties of the meta-atom comprised in the artificial surface \cite{9718037,9975205,9580737}. SP-EMSs are pre-configured, thus we design the wideband omni-directional retro-reflector by combining multiple, differently configured modules. The design of the single SP-EMS module is validated with full-wave simulations with high-frequency structure simulation (Ansys HFSS).
\item We evaluate the CRB on joint vehicle position estimation and RIS phase configuration uncertainty, for both frequency-dependent (wideband) and frequency-flat (narrowband) RIS behavior. RIS phase configuration is considered as deterministic (CRB) and random, a-priori information brought by coarse positioning at the vehicle side (hybrid CRB (HCRB)). The results highlight the importance of a wideband RIS modeling, as its frequency-dependent behavior act as a filter on the spectrum of the impinging signal. In some cases, a severe position-dependent filtering on the impinging signal contributes to decrease the CRB compared to the narrowband RIS behavior. Localization performance are compared against the case of a bare vehicle, showing the promising performance gap of using a EMS. 
\end{itemize}

\textit{Organization}: The paper is organized as follows: Section \ref{sect:ProblemFormulation} shows the experimental results aimed at motivating the usage of vehicle-lodged EMSs for localization, Section \ref{sect:system_model} outlines the system model, Sections \ref{sect:RIS_reflector_design} and \ref{sect:passive_IRS_reflector_design} reports the design of RIS- and SP-EMS-based reflector, respectively, Section \ref{sect:results} shows the localization performance using RISs/SP-EMSs while Section \ref{sect:conclusion} draws the conclusions of this work.

\textit{Notation}: Bold upper and lower-case letters stand for matrices and column vectors, respectively. $\left[\mathbf{{A}}\right]_{(i,j)}$ denotes the $(i,j)$ entry of matrix $\mathbf{A}$. Matrix transposition and conjugate transposition of $\mathbf{A}$ are indicated as $\mathbf{A}^T$ and $\mathbf{A}^H$, respectively. $\mathbb{E}$ denotes the expectation operator. With $\mathbf{a}\sim\mathcal{CN}(\boldsymbol{\mu},\mathbf{C})$ we denote a circularly complex multi-variate Gaussian random variable with mean $\boldsymbol{\mu}$ and covariance matrix $\mathbf{C}$. With $a\sim \mathcal{U}[b,c]$ we denote a uniformly distributed random variable between $b$ and $c$. $\mathbb{R}$ and $\mathbb{C}$ denote, respectively, the set of real and complex numbers. $\delta_n$ is the Kronecker delta.

\section{Problem Formulation and Methodology}\label{sect:ProblemFormulation}

\begin{figure}[b!]
    \centering
    \includegraphics[width=0.8\columnwidth]{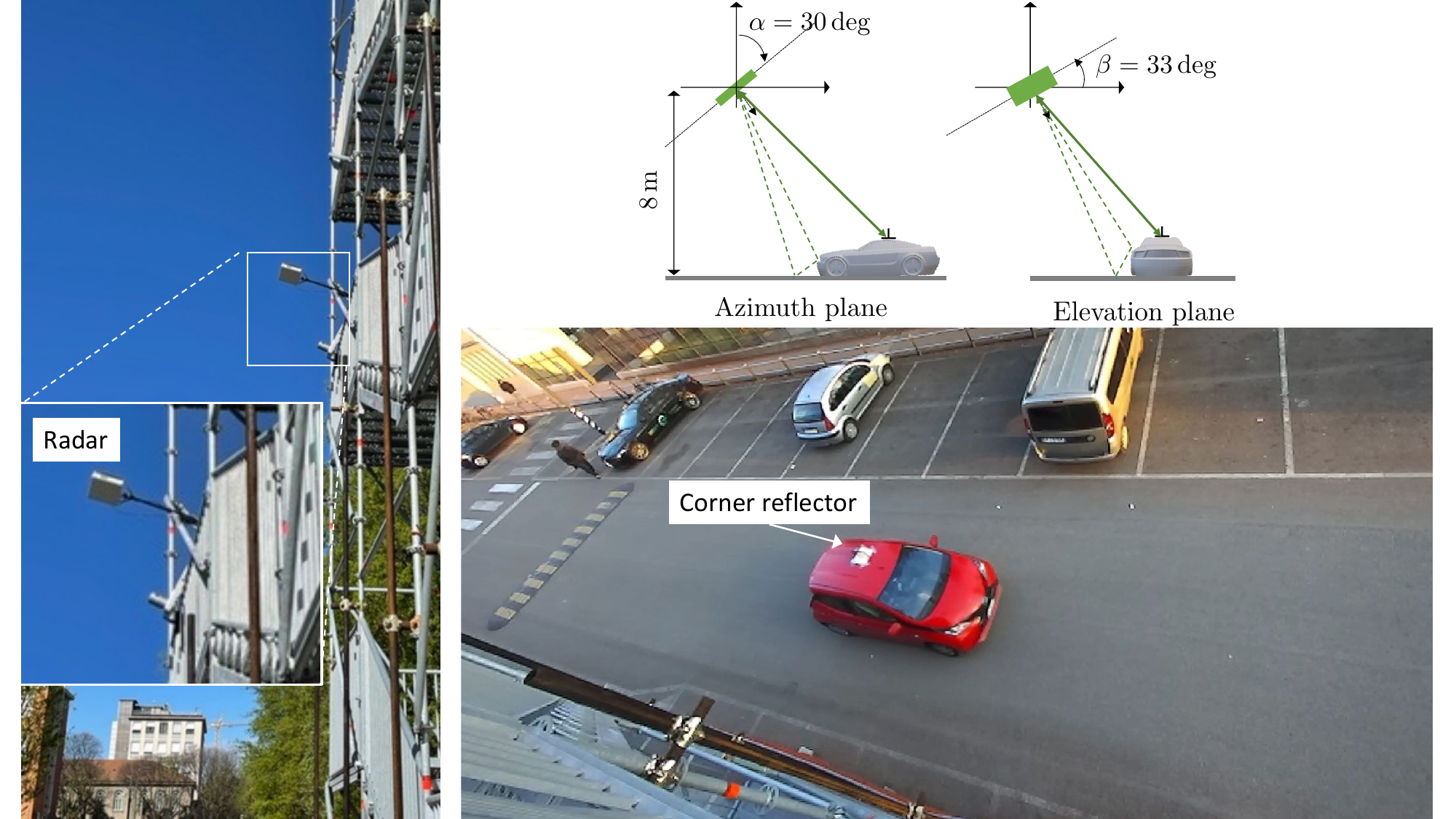}
    \caption{Experimental setup: a MIMO radar located at $8$ m height illuminates the street below where a car is purposely equipped with a corner reflector  }
    \label{fig:setup}
\end{figure}

Localizing the vehicle from a generic sensing terminal means estimating its position and possibly orientation based on the back-scattered signal. When the sensing terminal is located at the communication infrastructure (e.g., ISAC base stations (BSs)  or MIMO radar), its height compared to the one of the target is large, typically $8-25$ m for a BS~\cite{14rel} and $1.5-1.7$ m for a generic vehicle. When vehicles are illuminated from above, they back-scatter the incident radio signal depending on their geometric shape and geographic location w.r.t. the sensing terminal. The resulting sensing images for these extended targets typically exhibit highly distorted characters, which inhibit subjective interpretation and preclude quantitative analysis of spatial relationships. To gain insight of these effects, we report an example from an extensive radar data acquisition campaign mimicking a radar-/ISAC- BS imaging the surrounding. The geometry of the acquisition is shown in Fig. \ref{fig:setup}, where a MIMO radar operating at $77$ GHz frequency, with bandwidth $B=2.5$ GHz (range resolution of $6$ cm) and azimuth resolution of $1.4$ deg is located at $8$ m height above the road and illuminates the street below, where a car is travelling along a straight path at constant velocity. The true trajectory of the car for comparison is obtained using a real-time kinematic global positioning system (RTK-GPS) placed on the car roof.

\begin{figure}[t!]
    \centering
    \subfloat[]{\includegraphics[width=0.5\columnwidth]{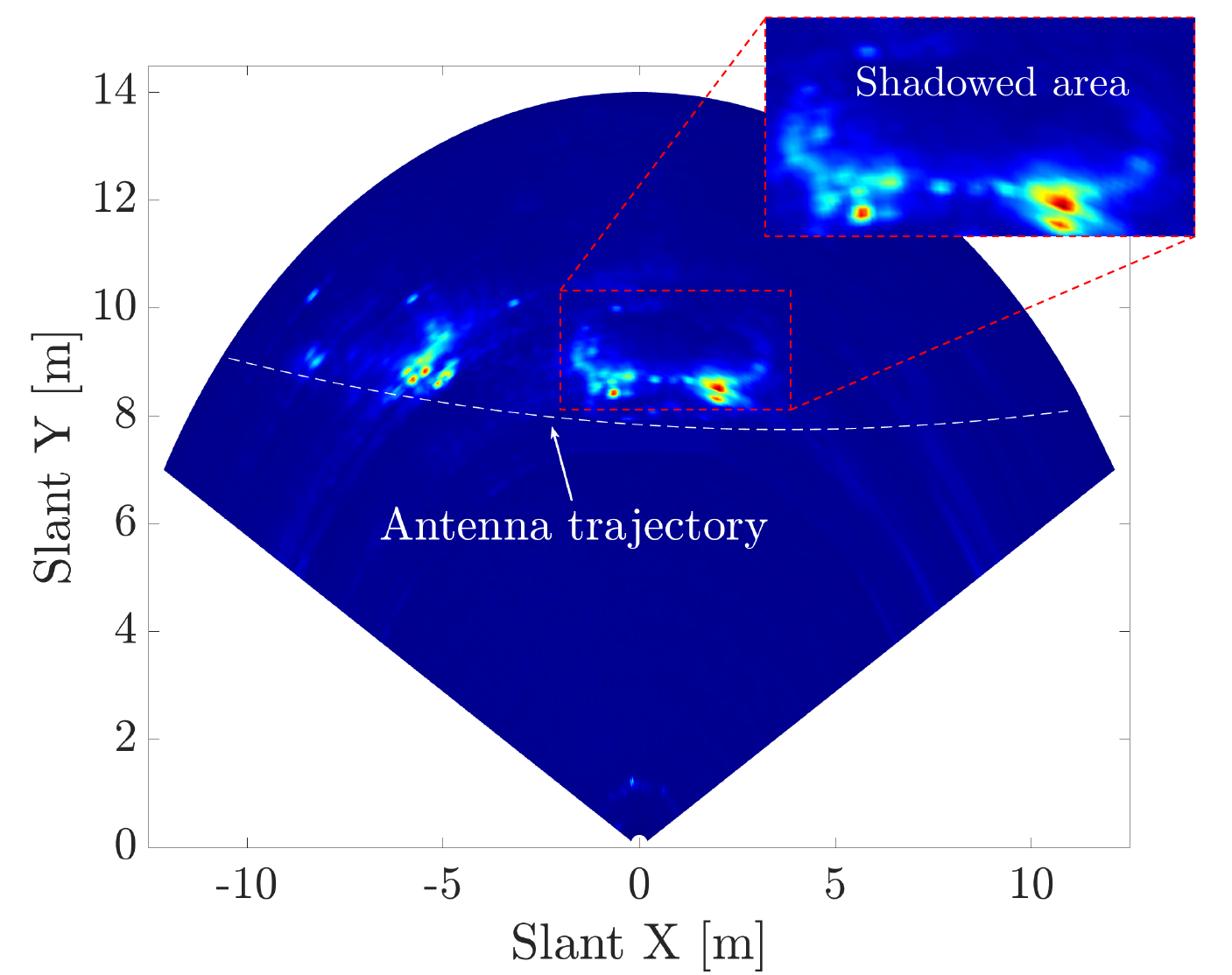}\label{fig:NoCorner}}
    \subfloat[]{\includegraphics[width=0.5\columnwidth]{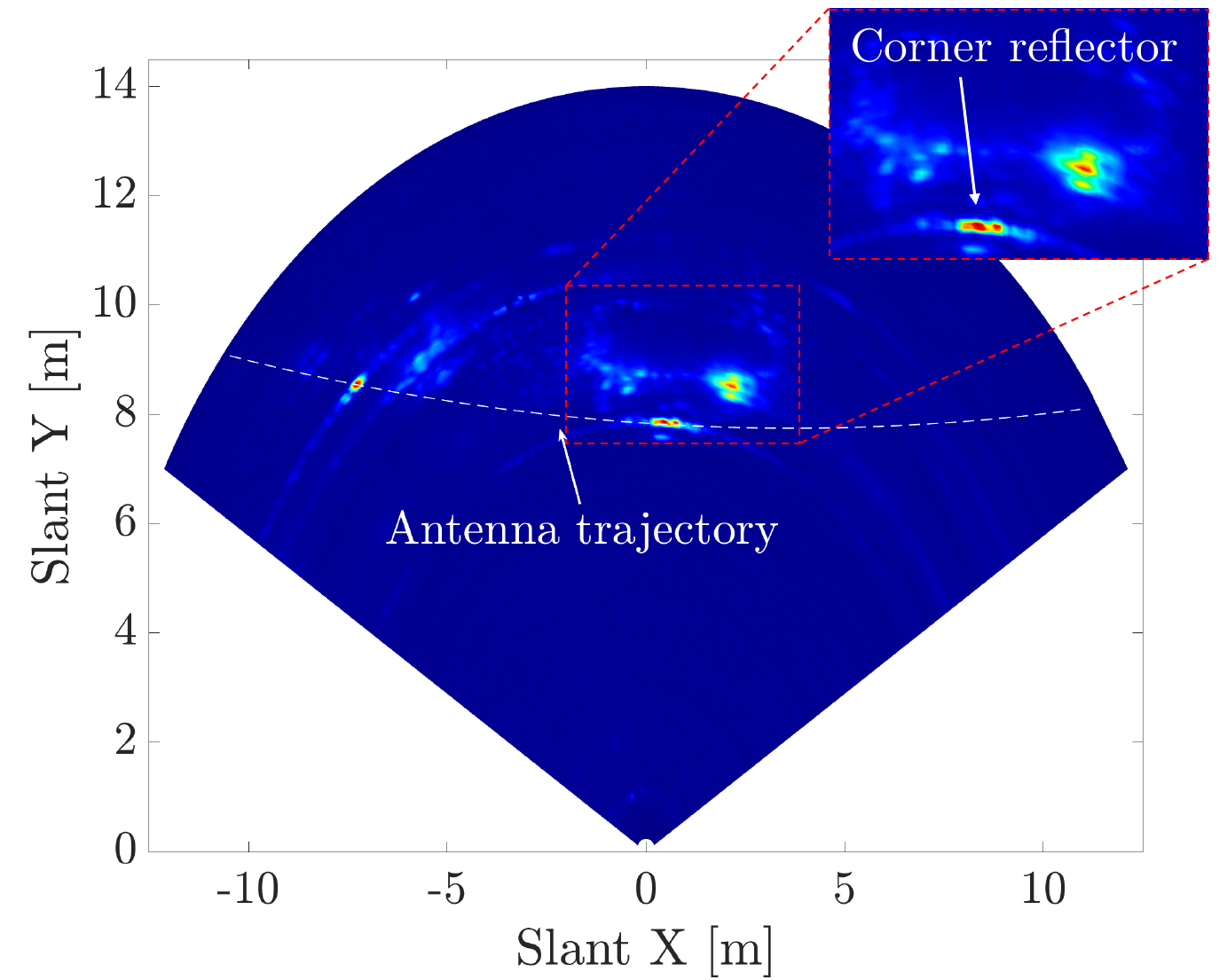}\label{fig:CarCorner}}
    \caption{Example of radar image of a car in slant range for (\ref{fig:NoCorner}) bare car and (\ref{fig:CarCorner}) car equipped with a corner reflector.}
    \label{fig:radar_image_car}
\end{figure}


The radar image in Fig. \ref{fig:NoCorner} shows two superimposed frames, with the vehicle being in a lateral position and subsequently in a central one (below the radar). The white dashed line is the RTK-measured true trajectory of the car, herein identified as the position of a possible mmWave/sub-THz antenna (notice that linear trajectory is curved in the slant-range plane). One can observe that the vehicle in the central position (rightmost and closest frame) reflects only with the lateral side, predominantly from the wheels and the alloy rims. A significant contribution to the back-scatter comes from double bouncing rays on the asphalt. The zoomed picture in Fig. \ref{fig:radar_image_car} shows a shadowed area due to the vehicle's silhouette (objects in this area are not visible to the radar). Instead, when the vehicle is in the lateral position, it reflects with the frontal-lateral corner side. In addition, we observe that the wheels appear farther away than the roof. These effects are known in remote sensing as foreshortening and layover \cite{naraghi1983geometric}, requiring accurate a-priori information on both the geometry and the physical structure of the target to be compensated for. If we are interested in the localization of a single, preferred point of the vehicle, for instance, the location of the mmWave/sub-THz antenna, having such an extended target whose peak reflectivity shifts according to the specific position in space requires advanced yet complicated tracking algorithms operating on raw radar data \cite{Folster2005,8730493,9399297,5466116}.

To facilitate the localization of the vehicle, this paper proposes to mount a reflector on the roof of the connected vehicles. To illustrate the potential of the proposed idea, we equipped the car with a four-sided corner reflector at the center of the roof, each side with $10\times10$ cm$^2$ area. The Fig. \ref{fig:CarCorner} represents the superposition of two radar frames, and it outlines that the corner reflector is visible in the radar image allowing for a precise localization. In other words, the vehicle behaves like a \textit{point target} in the slant-range acquisition frame. Nevertheless, mounting a corner reflector on the car's roof is neither aesthetic nor practical (it may interfere with the vehicle's aerodynamics). Hence, we propose to use EMSs as planar omnidirectional reflectors (\textit{EM markers}), which can better adapt to the car's roof shape. In the following sections, the design of omnidirectional reflectors based on EMSs is discussed.

\begin{figure}[!b]
    \centering
    \includegraphics[width=0.6\columnwidth]{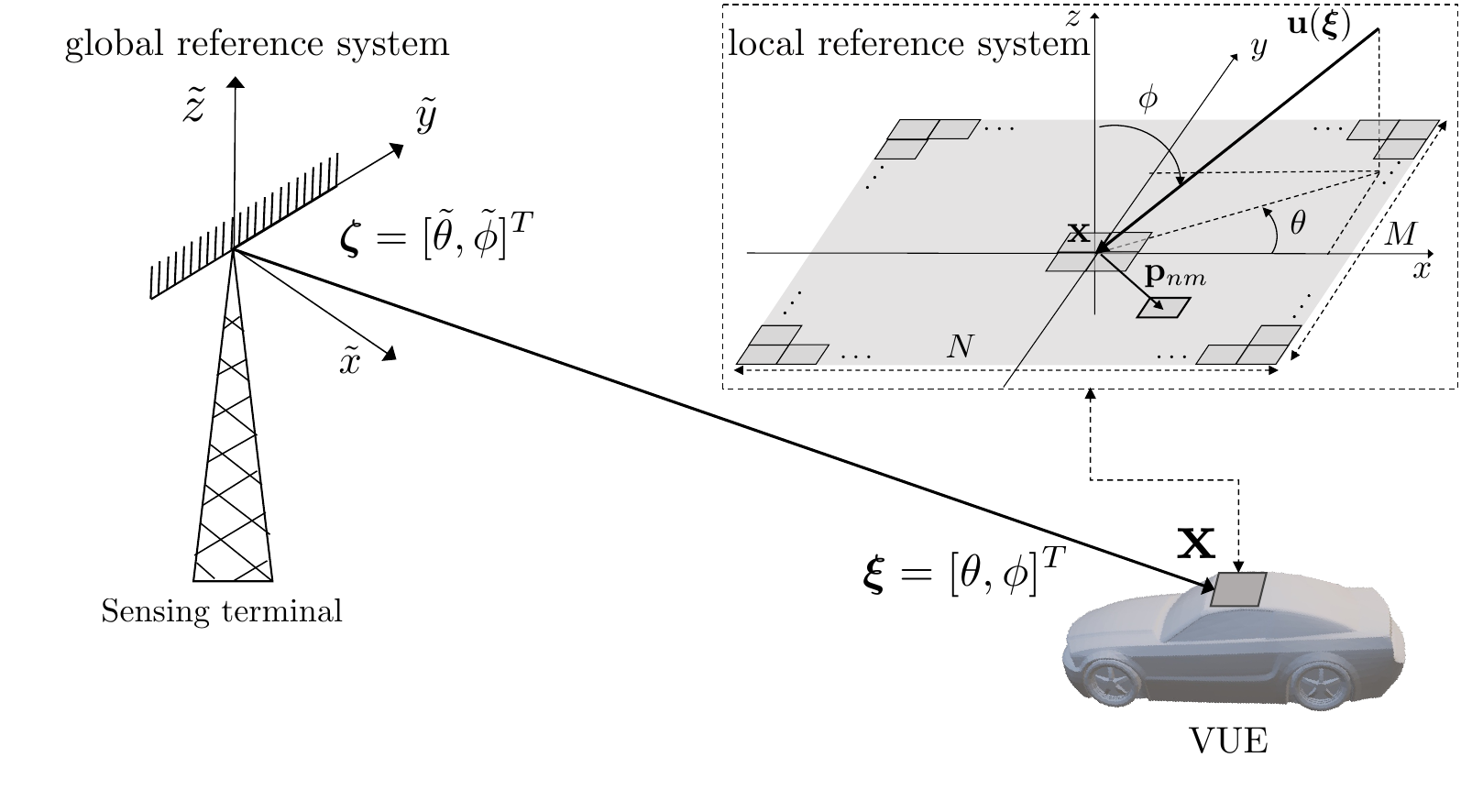}
    \caption{Global and EMS-local reference system}
    \label{fig:ref_system}
\end{figure}

\section{System Model}\label{sect:system_model}

Let us consider the vehicle mounting an EMS on the roof and the sensing terminal (e.g., an ISAC BS, a MIMO radar) that aims at localizing the vehicle via localizing the EMS. In the most general case, the sensing terminal implements $L$ measurement channels, whereas in the sensing jargon a measurement channel is a Tx-Rx pair. At each TX antenna, the sensing terminal emits the pass-band signal
\begin{equation}
    g(t) = s(t)e^{j2 \pi f_0 t}
\end{equation}
where the base-band signal $s(t)$ has bandwidth $B$ and the carrier frequency is $f_0$. Signal $s(t)$ can be intentionally (as for radar) or occasionally (as for ISAC) generated. We herein assume $s(t)$ is the same at each antenna, following the paradigm of a MIMO radar operating in time-division-multiplexing; the extension to antenna-specific signals is possible but not covered here. 
The phase center of the sensing terminal in the origin of the \textit{global} reference system, where the 3D positions of the transmitting (Tx) and receiving (Rx) antennas of the $\ell$-th channel are denoted as $\mathbf{s}_\ell\in\mathbb{R}^{3\times 1}$ and $\mathbf{r}_\ell\in\mathbb{R}^{3\times 1}$, respectively. 
The EMS is in the generic position $\mathbf{x} = [x,y,z]^\mathrm{T}\in\mathbb{R}^{3\times 1}$ and it is made by $N \times M$ elements along the $x$ and $y$ directions of the \textit{local} reference system, whose axes are rotated w.r.t. the global ones of the vehicle's heading $\psi$. Therefore, the position of the $(n,m)$-th element in global coordinates is  
\begin{align}\label{eq:position_RIS_el}
    \mathbf{x}_{nm} & = \mathbf{x} +  \mathbf{Q}_z(\psi)\mathbf{p}_{nm} 
\end{align}
for $n=-N/2,...,N/2-1$, $m=-M/2,...,M/2-1$, where $\mathbf{p}_{nm} = \left[nd, \,md,\,0\right]^T$ is the local position of the $(n,m)$-th element in local coordinates (for inter-element spacing $d$), matrix $\mathbf{Q}_z(\psi)\in \mathrm{SO}(3) \overset{\triangle}{=}\left\{\mathbf{Q}| \mathrm{det}(\mathbf{Q})=1, \mathbf{Q}\mathbf{Q}^T = \mathbf{I}_3\right\}$ defines the relative counterclockwise rotation of the local reference system around $z$ axis by angle $\psi$:
\begin{equation}
    \mathbf{Q}_{z}(\psi) = \begin{bmatrix}
        \cos\psi & -\sin\psi & 0\\
        \sin\psi & \cos\psi & 0 \\
        0&0&1
    \end{bmatrix}.
\end{equation}

\subsection{System model in time domain}

The generic expression of the received signal at the $\ell$-th measurement channel of the sensing terminal can be expressed as
\begin{equation}\label{eq:Rx_signal}
    \begin{split}
        y_\ell(t) & = \rho \sum_{n,m} e^{-j 2 \pi f_0 (\tau^i_{\ell,nm} + \tau^o_{\ell,nm})} e^{j\Phi_{nm}}\, s\left(t\hspace{-0.1cm}-\hspace{-0.1cm}\tau^i_{\ell,nm} \hspace{-0.1cm}- \hspace{-0.1cm}\tau^o_{\ell,nm}\right) + z_\ell(t) \overset{(a)}{\approx } \\
        & \overset{(a)}{\approx } \rho \, e^{-j 2 \pi f_0 (2\tau_0 + \Delta\tau^i_{\ell} + \Delta\tau^o_{\ell})} \sum_{n,m}
        e^{-j 4 \pi f_0 \Delta\tau_{nm}} e^{j\Phi_{nm}} s\left(t\hspace{-0.1cm}- \hspace{-0.1cm}2 \tau_0 \hspace{-0.1cm}- \hspace{-0.1cm}\Delta\tau^i_{\ell} \hspace{-0.1cm}- \hspace{-0.1cm}\Delta\tau^o_{\ell} \hspace{-0.1cm}- \hspace{-0.1cm}2 \Delta\tau_{nm}\right) + z_\ell(t), 
    \end{split}
\end{equation}
where $\rho$ denotes geometrical energy losses, the phase applied at the $(n,m)$-th element of the EMS is $\Phi_{nm}$ and $z_\ell(t)\in\mathcal{CN}(0,\sigma_z^2\delta_{\ell-k}\delta(t))$ is the additive white noise, uncorrelated over different channels.

The first, exact expression in \eqref{eq:Rx_signal} describes the most general case in which the wavefront across the EMS is non-planar, thus the EMS is in near-field w.r.t. the sensing terminal, and the same condition applies to the retro-reflected signal at the sensing terminal. Here, we account for the absolute delays between the $\ell$-th channel (Tx and Rx) and each element of the EMS, namely 
\begin{equation}\label{eq:delays}
    \tau^{i}_{\ell,nm} = \frac{ \|\mathbf{x}_{nm}-\mathbf{s}_\ell\|}{c}, \,\,\,\,\, \tau^o_{\ell,nm} = \frac{\|\mathbf{r}_\ell - \mathbf{x}_{nm}\|}{c}.
\end{equation}
This modeling is general and applies to any near-field scenario but often can be simplified by constraining the geometry of the problem. By assuming that the size of the EMS compared to the propagation distance is small, the incidence and reflected wavefront at both the sensing terminal and the EMS can be approximated as planar, and the delays can be linearized as follows:
\begin{align}
    \tau^{i}_{\ell,nm} &\simeq \underbrace{\frac{\|\mathbf{x}\|}{c}}_{\tau_0} + \underbrace{\frac{\mathbf{s}_\ell^T \mathbf{u}(\boldsymbol{\zeta})}{c}}_{\Delta \tau^i_\ell} + \underbrace{\frac{\mathbf{p}^T_{nm}\mathbf{u}(\boldsymbol{\xi}) }{c}}_{\Delta \tau_{nm}(\boldsymbol{\xi})} \label{eq:delay_inc_approx}\\
    \tau^{o}_{\ell,nm} &\simeq \underbrace{\frac{\|\mathbf{x}\|}{c}}_{\tau_0} + \underbrace{\frac{\mathbf{r}_\ell^T \mathbf{u}(\boldsymbol{\zeta})}{c}}_{\Delta \tau^o_\ell} + \underbrace{\frac{\mathbf{p}^T_{nm}\mathbf{u}(\boldsymbol{\xi}) }{c}}_{\Delta \tau_{nm}(\boldsymbol{\xi})},\label{eq:delay_out_approx}
\end{align}
where $\tau_0$ is the one-way propagation delay from the phase center of the sensing terminal to the phase center of the EMS, $\Delta \tau^i_\ell$, and $\Delta \tau^o_\ell$ are the residual (excess) delays at the sensing terminal w.r.t. to its phase center and, similarly, $\Delta \tau_{nm}(\boldsymbol{\xi})$ is the excess propagation delay at the EMS. In \eqref{eq:delay_inc_approx}-\eqref{eq:delay_out_approx}, angles \begin{align}
    \boldsymbol{\xi} &= [\theta, \phi]^T = J\left(-\mathbf{Q}_z(\psi) \mathbf{x}\right)\label{eq:RIS_angles}\\
    \boldsymbol{\zeta} &= [\tilde{\theta}, \tilde{\phi}]^T = J\left( \mathbf{x}\right)\label{eq:BS_angles}
\end{align}
are, respectively, the incidence/reflection angles at the EMS and the transmission/reception pointing angles to/from the EMS at the sensing terminal, obtained with the non-linear transformation
\begin{equation}
    J(\mathbf{x}): \begin{dcases}
    \theta = \arctan\left(\frac{y}{x}\right),\\
    \phi = \arccos\left(\frac{z}{\|\mathbf{x}\|}\right),
    \end{dcases}
\end{equation}
between global Cartesian coordinates to EMS-local spherical coordinates, while
\begin{align}
    \mathbf{u}(\boldsymbol{\xi}) = \left[\sin\phi \cos\theta, \sin\phi \sin\theta, \cos\phi\right]^T
\end{align}
is a unit vector defining a direction in local or global coordinates.
For the considered planar EMS, we have
\begin{equation}
    \Delta \tau_{nm}(\boldsymbol{\xi}) = \frac{d}{c}\left(n \sin\phi \cos\theta + m \sin\phi \sin\theta\right)
\end{equation}
and the phase at the EMS can be set according to a pair of incidence/reflection angles $\overline{\boldsymbol{\xi}}$ at center bandwidth ($f_0$) as:
\begin{equation}
    \Phi_{nm}(\overline{\boldsymbol{\xi}}) =  \frac{4\pi f_0 }{c} d \left(n \sin\overline{\phi} \cos\overline{\theta} + m \sin\overline{\phi} \sin\overline{\theta}\right)
\end{equation}
so as to compensate the propagation phase $4 \pi f_0 \Delta \tau_{nm}$ and maximize the reflection gain at $f_0$ for $\overline{\boldsymbol{\xi}}=\boldsymbol{\xi}$.

\subsection{System model in frequency domain}

When the bandwidth $B$ of the Tx signal is large enough to induce spatial wideband effects at the EMS and/or at the sensing terminal, the reflection gain is maximized only around the carrier frequency $f_0$, and \textit{reflection beam squinting} effects arise. Spatial wideband effect is a well-known issue of radars \cite{8443598}, where space-time adaptive processing techniques are employed to deal with antenna-specific delays on the base-band pulse. 
The system model for far-field, i.e,, under approximation $(a)$ in \eqref{eq:Rx_signal}, is conveniently written in the frequency domain by operating a Fourier transform over $y_\ell(t)$, obtaining 
 \begin{equation}\label{eq:Rx_signal_frequency}
 \begin{split}
      Y_\ell(f) & \overset{(a)}{\approx} \rho \, e^{-j 2 \pi (f_0+f) (2\tau_0 + \Delta\tau^i_{\ell} + \Delta\tau^o_{\ell})} \underbrace{\sum_{n,m} e^{-j 4 \pi (f_0+f) \Delta\tau_{nm} } e^{j\Phi_{nm}(\overline{\boldsymbol{\xi}})}}_{\beta(f,\gamma,\boldsymbol{\xi}\lvert \overline{\boldsymbol{\xi}})} S(f) + Z_\ell(f)
 \end{split}
 \end{equation}
where $f$ denotes the base-band frequency (i.e., around $f_0$), $\beta(f,\boldsymbol{\xi}\lvert \overline{\boldsymbol{\xi}})$ is the frequency-dependent scattering amplitude term incorporating both the geometrical energy losses ($\rho$) and the reflection gain of the RIS, while $Z_\ell(f)\sim\mathcal{CN}(0,N_0\delta(f)\delta_{\ell-k})$ is the noise in frequency, with power spectral density $N_0$, uncorrelated across channels. Model \eqref{eq:Rx_signal_frequency} applies to those scenarios where the bandwidth $B$ of the impinging signal $s(t)$ is so large that the RCS of the EMS depends on the base-band frequency $f$. The model of $\beta(f,\gamma,\boldsymbol{\xi}\lvert \overline{\boldsymbol{\xi}})$ follows from the radar equation \cite{ulaby2014microwave}:
\begin{equation}
    \beta(f,\gamma,\boldsymbol{\xi}\lvert \overline{\boldsymbol{\xi}}) = \sqrt{\frac{ c^2}{f_0^2 (4\pi)^3 \|\mathbf{x}\|^4} \Gamma_\mathrm{ris}(f,\boldsymbol{\xi}|\overline{\boldsymbol{\xi}})} \;e^{j \gamma}
\end{equation}
where $\Gamma_\mathrm{ris}(f,\boldsymbol{\xi}|\overline{\boldsymbol{\xi}})$ is the RCS of the EMS and $\gamma$ is a phase term modeling residual phase uncertainties about the target (the EMS) arising from Tx-Rx circuitry, Doppler effects from motion, etc. Notice that the RCS of the EMS is an implicit function of the EMS position $\mathbf{x}$ via angles $\boldsymbol{\xi}$.

\section{RIS-Based Reflector Design}\label{sect:RIS_reflector_design}


Let us consider the usage of a RIS as a retro-reflector. If the vehicle can configure the phase of the RIS with the ideal (true) angles $\overline{\boldsymbol{\xi}}=\boldsymbol{\xi}$, we have:
\begin{equation}
    \Phi_{nm}(\boldsymbol{\xi}) =   \frac{4\pi f_0 }{c} d \left(n \sin\phi \cos\theta + m \sin\phi  \sin\theta\right).
\end{equation}
In this setting, the sensing system experiences the maximum possible RCS of the RIS at $f_0$, that be generally expressed as follows
\begin{equation}\label{eq:RCS}
    \Gamma_\mathrm{ris}(f,\boldsymbol{\xi}|\overline{\boldsymbol{\xi}}) = \Gamma^\mathrm{max}_\mathrm{ris}(f) \; G(f,\boldsymbol{\xi}|\overline{\boldsymbol{\xi}})
\end{equation}
where $\Gamma^\mathrm{max}_\mathrm{ris}(f)$ is the peak RCS of the RIS, function of the frequency and $G(f,\boldsymbol{\xi}|\overline{\boldsymbol{\xi}})$ is the normalized array factor of the RIS. The expression of $G(f,\boldsymbol{\xi}|\overline{\boldsymbol{\xi}})$ is 
\begin{equation}\label{eq:RIS_array_factor}
        \begin{split}
            G(f,\boldsymbol{\xi}|\overline{\boldsymbol{\xi}}) & = \frac{1}{N^2M^2}\bigg\lvert\sum_{n,m}  e^{-j 4 \pi (f_0+f) \Delta\tau_{nm}(\boldsymbol{\xi}) } e^{j\Phi_{nm}(\overline{\boldsymbol{\xi}})}\bigg\rvert^2= \\
             &= \frac{1}{N^2M^2}\bigg\lvert \frac{\sin\left( N \alpha_x(f,\boldsymbol{\xi}|\overline{\boldsymbol{\xi}}) \right)}{\sin\left(\alpha_x(f,\boldsymbol{\xi}|\overline{\boldsymbol{\xi}})\right)}
            \frac{\sin\left(M \alpha_y(f,\boldsymbol{\xi}|\overline{\boldsymbol{\xi}})\right)}{\sin\left(\alpha_y(f,\boldsymbol{\xi}|\overline{\boldsymbol{\xi}})\right)}\bigg\rvert^2 
        \end{split}
    \end{equation}
where
\begin{align}
\alpha_x(f,\boldsymbol{\xi}|\overline{\boldsymbol{\xi}}) & = \frac{2\pi d}{c}\left[f_0 \sin\overline{\phi}\cos\overline{\theta} - (f_0+f)\sin\phi\cos\theta\right],\\
\alpha_y(f,\boldsymbol{\xi}|\overline{\boldsymbol{\xi}}) & = \frac{2\pi d}{c}\left[f_0 \sin\overline{\phi}\sin\overline{\theta} - (f_0+f)\sin\phi\sin\theta\right].
\end{align}
For $\overline{\boldsymbol{\xi}}=\boldsymbol{\xi}$, the array factor is $\leq 1$ except for $f=0$ (i.e., at center bandwidth $f_0$) due to the frequency-dependent squinting effects.
The peak RCS can be well approximated by the RCS of a metallic plate of area $A_\mathrm{ris}=NMd^2$ perpendicular to the incidence/reflection direction as \cite{Bjornson2020, Ellingson2021}:
\begin{equation}\label{eq:RCS_max}
    \Gamma^\mathrm{max}_\mathrm{ris}(f) = \frac{4 \pi f^2 A^2_\mathrm{ris}}{c^2} = \frac{4 \pi f^2 (NM)^2 d^4}{c^2} \quad \text{[m$^2$]}.
\end{equation}
\begin{figure}[!t]
    \centering
    \subfloat[][$A_\mathrm{ris}=5\times 5$ cm$^2$]{\includegraphics[width=0.5\columnwidth]{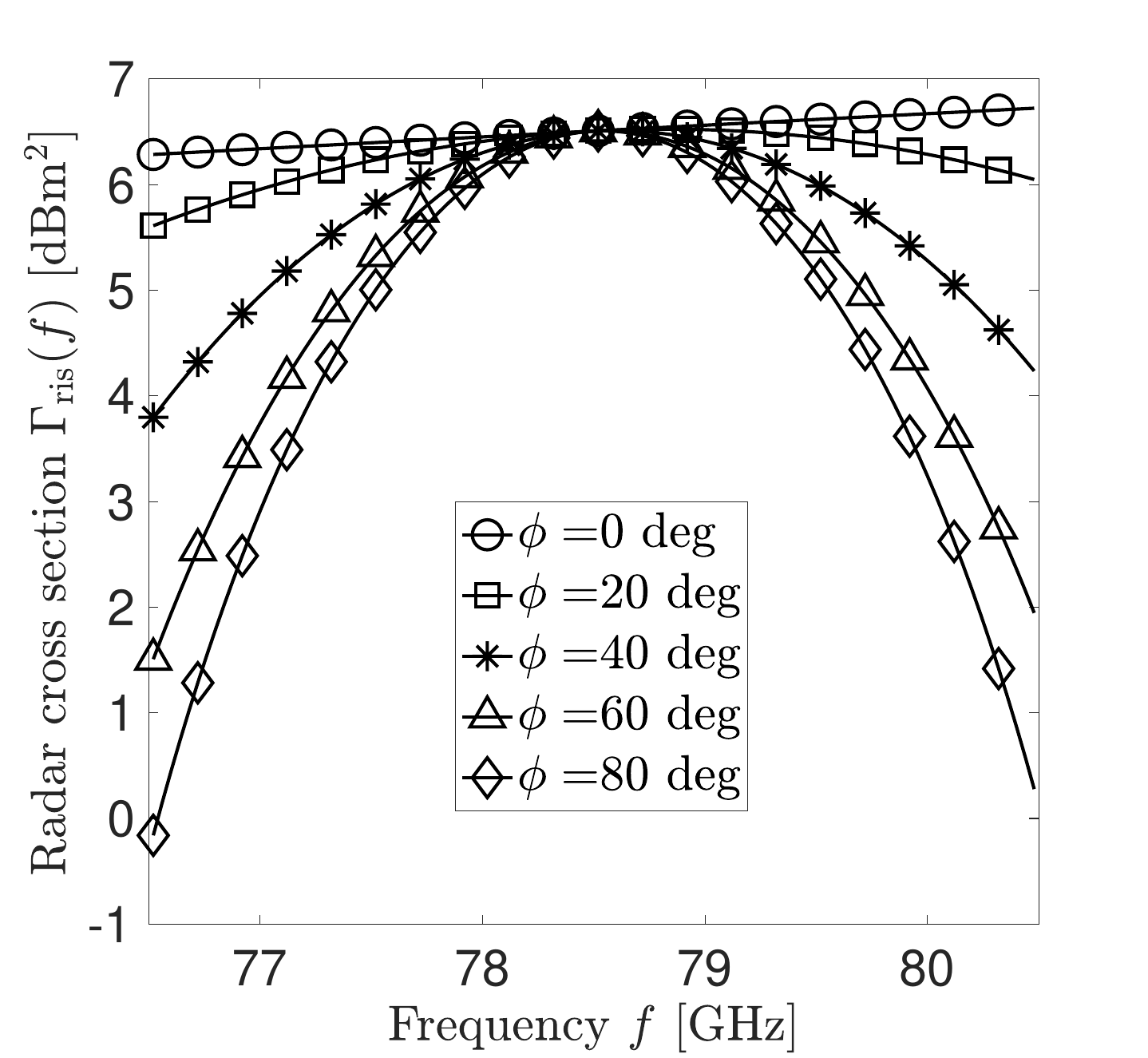}\label{subfig:RCS_RIS_5x5}}
    \subfloat[][$A_\mathrm{ris}=10\times 10$ cm$^2$]{\includegraphics[width=0.5\columnwidth]{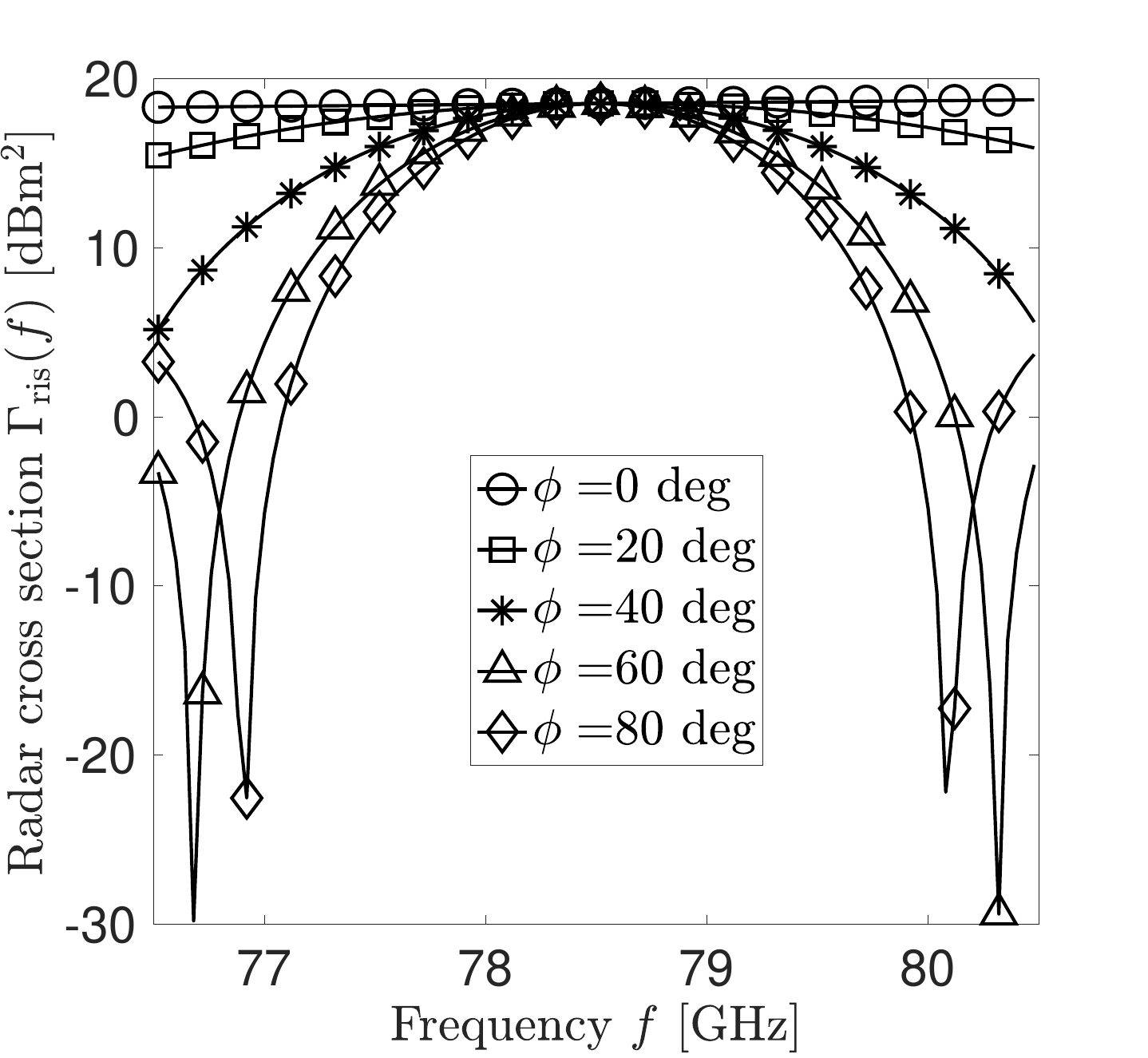}\label{subfig:RCS_RIS_10x10}}
    \caption{RCS of the RIS varying the incident signal frequency and elevation angle $\phi$, for (\ref{subfig:RCS_RIS_5x5}) a $A_\mathrm{ris}=5\times 5$ cm$^2$ RIS and (\ref{subfig:RCS_RIS_10x10}) a $A_\mathrm{ris}=10\times 10$ cm$^2$ RIS. }
    \label{fig:RCS_vs_frequency}
\end{figure}
%
%
%
For a $10\times 10$ cm$^2$ RIS operating at $f_0=77$ GHz frequency, the RCS@$f_0$ is $72.5$ m$^2$, comparable with the one of a tetrahedral corner reflector of $10$ cm for each side. The RIS size, namely the number of elements $N\times M$ shall be designed to achieve 
\begin{equation}\label{eq:rcsConstraint}
    \Gamma^\mathrm{max}_\mathrm{ris}(f_0) \geq \Gamma_\mathrm{min}
\end{equation}
where $\Gamma_\mathrm{min}$ is a minimum RCS that ensures the detectability of the RIS. A rule of thumb, here proposed, is to compute it based on the ratio between the corner reflector's RCS and the maximum measured car's RCS. Based on the experimental radar measurements reported in Section \ref{sect:ProblemFormulation}, we observed a ratio of $10$ dB. The RCS of the corner reflector $\Gamma_\mathrm{CR}$ with side $a$ is defined as \cite{ulaby2014microwave}
\begin{equation}\label{eq:RCS_CR}
    \Gamma^\mathrm{max}_\mathrm{CR}(f_0)  = \frac{12\pi f_0^2 a^4}{c^2}, \quad \text{[m$^2$]}.
\end{equation}
The area of the RIS can be computed based on \eqref{eq:RCS_max} and \eqref{eq:RCS_CR} as
\begin{equation}\label{eq:RCS_RIS_criterion}
    \Gamma^\mathrm{max}_\mathrm{ris}(f_0) = \frac{\Gamma^\mathrm{max}_\mathrm{CR}(f_0)}{10} \longrightarrow A_\mathrm{ris}\approx a^2 \sqrt{\frac{3}{10}}
\end{equation}
corresponding to a RIS with an area $A_\mathrm{ris} = 7.4 \times 7.4$ cm$^2$. However, even in the ideal case of a perfect phase configuration, the the beam squinting cannot be ignored in the experienced RCS by a sensing system illuminating a RIS. Fig. \ref{fig:RCS_vs_frequency} shows the trend of the RCS $\Gamma_\mathrm{ris}(f,\boldsymbol{\xi}|\boldsymbol{\xi})$ as function of the frequency, for $\theta=0$ deg (azimuth) and $\phi=0,20,40,60,80$ deg (elevation). It is immediate to notice that, for increasing grazing angles $\phi$, the frequency-selectivity of the RIS increases as well, \textit{filtering} the impinging signal. For a typical bandwidth $B=4$ GHz, $\phi=60$ deg, the RCS decreases by $4$ dB for a $5\times 5$ cm$^2$ RIS and by $30$ dB for a $10\times 10$ cm$^2$ RIS.

\subsection{RIS configuration in realistic settings}\label{subsect:posErr}

\begin{figure}[!t]
    \centering
    \includegraphics[width=0.5\columnwidth]{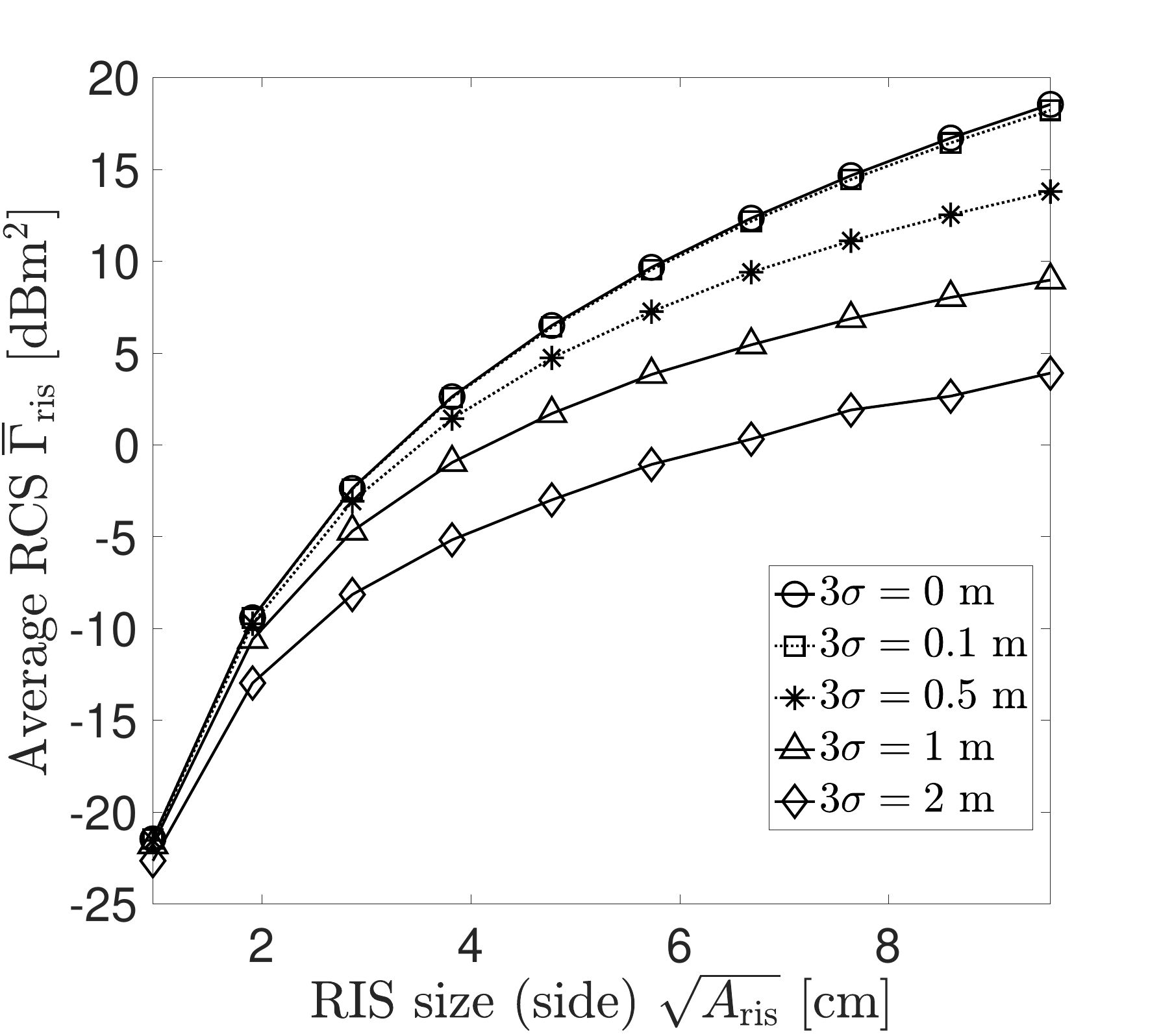}
    \caption{Effect of positioning error on the experienced RCS: average RCS $\overline{\Gamma}_\mathrm{ris}(\boldsymbol{\xi})$ (in dBm$^2$) as function of the RIS size varying the positioning error $\sigma$.}
    \label{fig:RCS_vs_pos}
\end{figure}

In practice, the vehicle does not have the perfect knowledge of the RIS pointing angles $\boldsymbol{\xi}$, and the RIS phase configuration leverages on suitable estimated angles $\widehat{\boldsymbol{\xi}}$. A typical source of error on $\boldsymbol{\xi}$ is the mobility of the vehicle w.r.t. the sensing terminal, that makes any estimated data $\widehat{\boldsymbol{\xi}}$ rapidly outdated. The phase pattern applied to the RIS reflector is
\begin{equation}\label{eq:estimated_phase}
    \Phi_{nm}(\widehat{\boldsymbol{\xi}}) = \frac{4\pi f_0 }{c} d \left(n \sin\widehat{\phi} \cos\widehat{\theta} + m \sin\widehat{\phi}  \sin\widehat{\theta}\right) 
\end{equation}
and the corresponding RCS experienced by the sensing system can be computed using \eqref{eq:RCS} and plugging $\overline{\boldsymbol{\xi}}= \widehat{\boldsymbol{\xi}}$ in \eqref{eq:RIS_array_factor}. Notice that leveraging on-board positioning systems (e.g., GPS) is not generally sufficient to attain the desired localization performance as \textit{(i)} the spatial resolution of sensing systems may be much higher that the positioning accuracy of current technologies\footnote{A typical GPS has a positioning accuracy in the order of $1$ m to few m (without urban canyoning effects), while a MIMO radar working at $f_0=77$ GHz with $25$ cm of aperture ($\approx 100$ \textit{virtual} antennas) and $3$ GHz of bandwidth (e.g., \cite{TI_ref_MMWCAS}) has a resolution cell of $15\times 5$ cm at $20$ m distance, comparable with the one of an expensive RTK setup} and \textit{(ii)} the average experienced RCS is low. To gain insight on this latter aspect, let us consider the case in which the vehicle has a coarse estimate of the incidence/reflection angles
\begin{equation}\label{eq:directionWithErr}
    \widehat{\boldsymbol{\xi}} = [\widehat{\theta},\widehat{\varphi}]^T = J\left( -\mathbf{Q}_z(\psi) \widehat{\mathbf{x}}\right)
\end{equation}
where $\widehat{\mathbf{x}}\sim\mathcal{N}(\mathbf{x},\mathbf{C}_{\mathbf{x}})$ is the on-board position measurement made by the vehicle, characterized by covariance $\mathbf{C}_{\mathbf{x}} = \sigma^2 \mathbf{I}_3$\footnote{The choice of a circular Gaussian error model, as well as a diagonal covariance matrix, is dictated by the seek for simplicity in the numerical derivations herein, but does not limit the validity of the results.}.  Figure \ref{fig:RCS_vs_pos} shows the trend of the average RCS $\overline{\Gamma}_\mathrm{ris}(\boldsymbol{\xi})$ experienced by the sensing system varying the RIS size (assuming a squared RIS of $N\times N$ elements, spaced by $d=\lambda_0/4$ at $f_0 = 78.5$ GHz) when the true RIS position is $\mathbf{x}=[10,0,-6.5]^\mathrm{T}$ and it is known with variable accuracy $3\sigma = 0,\, 0.1,\, 0.5,\,1,\,2 $ m. The BS is in the origin of the reference system, at $6.5$ m above the RIS. Increasing the RIS size leads to a progressive gap between the expected maximum RCS $\Gamma^\mathrm{max}_\mathrm{ris}(f)$ and the average one $\overline{\Gamma}_\mathrm{ris}(f,\boldsymbol{\xi})$, degrading with $\sigma$. The consequence of an inaccurate estimation of $\widehat{\boldsymbol{\xi}}$ is that the RIS will not be configured optimally, resulting in a loss of the experienced RCS, which is more severe as the spatial selectivity of the RIS increases. 

An effective solution to overcome the aforementioned issue is to refine the estimation of the RIS configuration angles $\boldsymbol{\xi}$ by a local search over the angular span defined by the confidence on the initial, coarse estimate $\widehat{\boldsymbol{\xi}}$. Let us define $\mathbf{C}_{\boldsymbol{\xi}}$ as the covariance matrix of the angular estimation error $\widehat{\boldsymbol{\xi}}-\boldsymbol{\xi}$. The optimal RIS configuration angles are obtained with the following \textit{RIS alignment procedure}:
\begin{equation}\label{eq:sweeping}
    \boldsymbol{\xi}_{opt}=\underset{\boldsymbol{\xi}_{k,q}\in \Xi_\mathrm{ris}}{\mathrm{argmax}} \;\Gamma_\mathrm{ris}(f_0,\boldsymbol{\xi}_{k,\ell}|\boldsymbol{\xi})\longrightarrow \left\{\Phi_{nm}(\boldsymbol{\xi})\right\}
\end{equation}
where $\Gamma_\mathrm{ris}(f_0,\boldsymbol{\xi}_{k,q}|\boldsymbol{\xi})$ is the RCS at $f_0$ experienced by choosing the reflection angles $\boldsymbol{\xi}_{k,q}= [\theta_k,\varphi_q]^T$ within the finite set $\Xi_\mathrm{ris} = \Theta_\mathrm{ris} \times \Psi_\mathrm{ris}$. Set $\Xi_\mathrm{ris}$ is \textit{dynamically} updated based on the estimated angles $\widehat{\boldsymbol{\xi}}$ and the related uncertainties $\sigma_{\theta} = \sqrt{\left[\mathbf{C}_{\boldsymbol{\xi}}\right]_{(1,1)}}$ and $\sigma_{\varphi} = \sqrt{\left[\mathbf{C}_{\boldsymbol{\xi}}\right]_{(2,2)}}$, as
\begin{align}
    \Theta_\mathrm{ris} & = \left\{ \theta_k \,\bigg\lvert\, \theta_k = \widehat{\theta} + k \left(\frac{\Delta\theta}{\kappa\sigma_\theta}\right), k = -\frac{K}{2},...\frac{K}{2}\right\} \label{eq:THETA}\\
    \Psi_\mathrm{ris} & = \left\{ \varphi_q \, \bigg\lvert\, \varphi_q = \widehat{\varphi} + q \left(\frac{\Delta\varphi}{\kappa\sigma_\varphi}\right), q = -\frac{Q}{2},...\frac{Q}{2}\right\}\label{eq:PSI}
\end{align}
where: \textit{(i)} $\Delta \theta$ and $\Delta \varphi$ are the reflection beamwidths in azimuth and elevation, \textit{(ii)} $K = \lceil 2\kappa\sigma_\theta/\Delta\theta \rfloor$ and $Q = \lceil 2\kappa\sigma_\varphi/\Delta\varphi \rfloor$ are the codebooks' cardinalities, and \textit{(iii)} $\kappa\geq1$ is a proper factor accounting for a pre-defined confidence interval, e.g., $\kappa=3$. Sets $\Theta_\mathrm{ris}$ and $\Psi_\mathrm{ris}$ denote the spanned azimuth and elevation angles, respectively, within a given confidence. The proposed method converges to the optimal value of configuration angles, with a residual error that depends on the codebook cardinality. In the following, we provide some considerations on the RIS alignment procedure related to mobility. 

\subsection{Mobility considerations}
The duration of the RIS alignment procedure must ensure that the optimal RIS configuration $\boldsymbol{\xi}_{opt} = \boldsymbol{\xi}$ does not change within the interval $T_{\mathrm{train}}$. For a vehicle is moving at a speed $v$ in the coverage area of the sensing terminal, it can be shown that  the training time shall fulfill the following inequality:
\begin{equation}\label{eq:training_time_upperbound}
    T_{\mathrm{train}} \leq T_{\mathrm{PRI}} \times  \underbrace{\mathrm{min}\left(\frac{\Delta\theta D_{\mathrm{min}} }{v T_{\mathrm{PRI}}},\frac{\Delta\varphi D_{\mathrm{min}}}{v T_{\mathrm{PRI}} \cos\varphi_{\mathrm{min}}} \right)}_{(K\times Q)_\mathrm{max}}
\end{equation}
where $T_{\mathrm{PRI}}$ is the fundamental unit of time taken to test a single reflection pattern, which does not depend on the RIS switching capability, but rather can be well approximated by the pulse repetition interval (PRI) of the sensing system (tens of $\mu$s~\cite{TI_ref_MMWCAS}), and the second term is the maximum number of tested reflection beams. The latter is a function of the vehicle velocity $v$, the minimum distance of the RIS from the sensing terminal $D_\mathrm{min}$, the minimum elevation incidence angle on the RIS $\varphi_{\mathrm{min}}$. Notice that the design of the RIS-alignment procedure does not need to explicitly take into account the Doppler shift of the vehicle in \eqref{eq:training_time_upperbound}, as the mobility is described by velocity $v$. In fact, Doppler shift does not affect the received signal power as for \eqref{eq:Rx_signal_frequency}.\\
\textit{Example}: Considering a typical urban scenario, where $v=50$ km/h, a typical PRI $T_{\mathrm{PRI}}=50$ $\mu$s, $f_0 = 77$ GHz, $D_{\mathrm{min}}=10$ m ($\varphi_{\mathrm{min}}=45$ deg), and $Nd=10$ cm ($\Delta\theta=0.5$ deg, $\Delta\varphi=2.5$ deg) we have that the maximum cardinality of $\Xi_\mathrm{ris}$ is $(K\times Q)_\mathrm{max} \leq 130$.
For a positioning uncertainty of $3\sigma = 2$ m, we obtain $(K\times Q)_\mathrm{max} = 60$, which is far below the urban scenario limit. However, in highway scenarios with $v=150$ km/h, the PRI of the system must be reduced for non-ambiguous velocity estimation (see \cite{skolnik}), e.g., $T_{\mathrm{PRI}}\leq 30$ $\mu$s, thus $K\times Q \leq 70$ is close to the upper limit. In these settings, a proper tracking of the reflection beam or, equivalently, of the position, allows reducing the uncertainty and the related codebook depth.

\section{SP-EMS-based Reflector Design }\label{sect:passive_IRS_reflector_design}

\begin{figure}[!t]
    \centering
    \includegraphics[width=0.5\columnwidth]{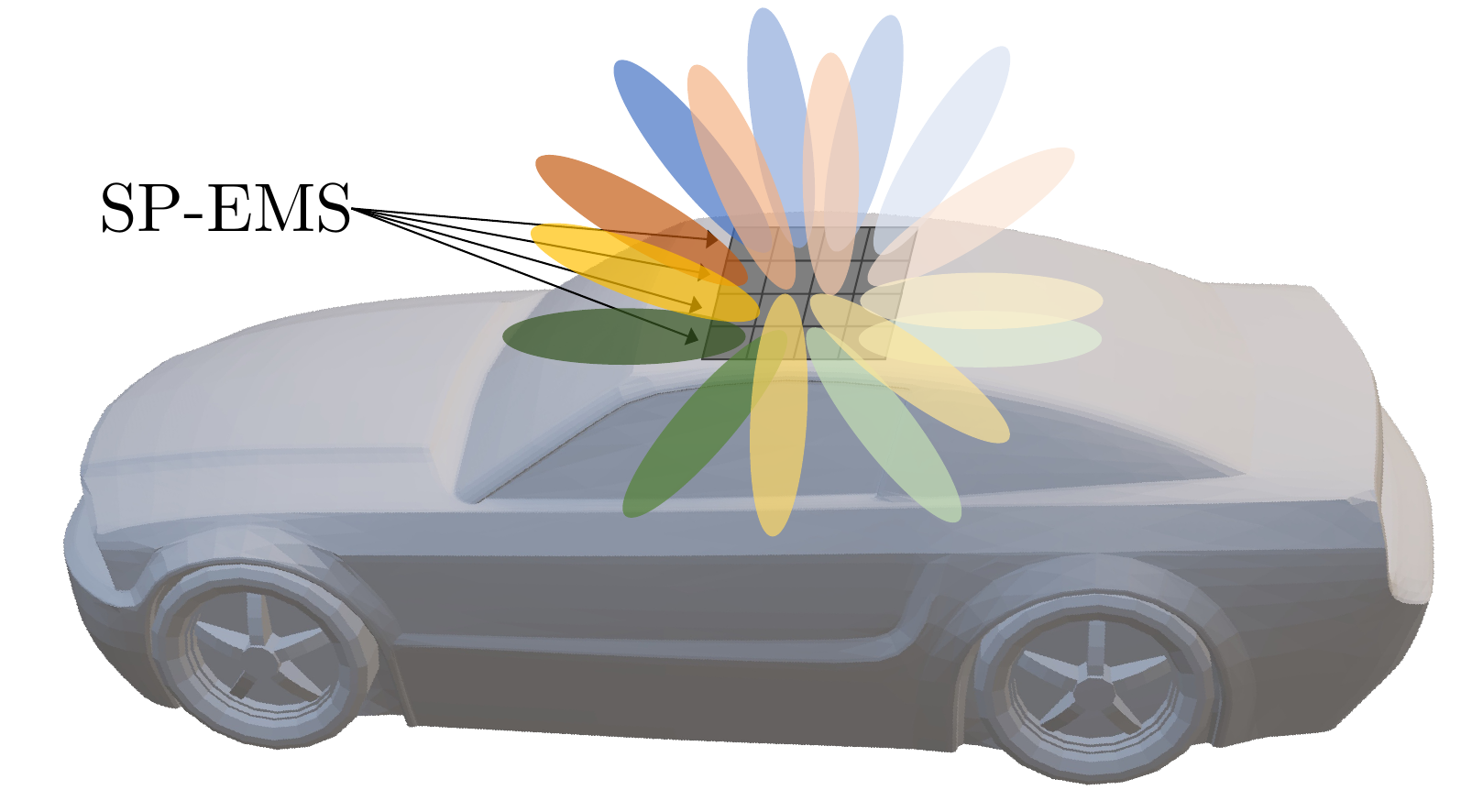}
    \caption{SP-EMSs on the roof of the vehicle for omnidirectional localization. Different colors denote retro-reflection from different modules. }
    \label{fig:idea_SPEMS}
\end{figure}

A cost-effective alternative to RIS-based reflectors, considered hereinafter, consists of a fully passive and static retro-reflector, belonging to the class of SP-EMSs \cite{9718037,9975205,9580737}. Since the wave manipulation capabilities of an SP-EMS are defined in the design phase, the maximum retro-reflection can be guaranteed only toward the designated direction. Hence, a suitable solution is to have multiple SP-EMSs, hereafter referred to as \textit{modules}, composing an omnidirectional retro-reflector, whereby each module is differently configured to retro-reflect the impinging signal from a specific direction, as indicated in Fig. \ref{fig:idea_SPEMS}. To this aim, consider the case of having $P$ modules. The location of the $(n,m)$-th meta-atom of the $p$-th module is therefore:
\begin{align}\label{eq:position_skin_el}
    \mathbf{x}_{p,nm} & = \mathbf{x}_p +  \mathbf{Q}_z(\psi)\mathbf{p}_{nm},
\end{align}
where $\mathbf{x}_p = \mathbf{x} + \mathbf{p}_p$ is the location of the $p$-th module in global coordinates ($\mathbf{p}_p$ is the position in local coordinates). The $p$-th module is configured to retro-reflect the impinging signal from direction $\overline{\boldsymbol{\xi}}_p=[\cos\overline{\theta}_p,\sin\overline{\phi}_p]^T$:
\begin{equation}\label{eq:estimated_phase_SPEMS}
    \Phi_{p,nm}(\overline{\boldsymbol{\xi}}_p) =   \frac{4\pi f_0 }{c} d \left(n \sin\overline{\phi}_p \cos\overline{\theta}_p + m \sin\overline{\phi}_p  \sin\overline{\theta}_p\right).
\end{equation}
The set of modules composing the SP-EMS can be designed such that: (1) the minimum RCS of each module is sufficient to make it detectable in the sensing image, (2) the retro-reflection main lobe of each module is well separated to the others in the angular domain, i.e., for a specific incidence direction $\overline{\boldsymbol{\xi}}$, only one module will effectively contribute to the retro-reflection and (3) each module must work over the whole sensing bandwidth, with no or limited filtering of the impinging signal (see Section \ref{sect:RIS_reflector_design}). 

Criterion (1) follows the empirical method derived in Section \ref{sect:RIS_reflector_design}, \eqref{eq:RCS_RIS_criterion}, providing module sizes ranging from $4\times4$ cm$^2$ to $10\times10$ cm$^2$. Criterion (2) allows to minimize the number of skins to be implemented in order to achieve full angular coverage, and it enables to approximate the RCS of the omnidirectional reflector as follows:
\begin{equation}\label{eq:RCS_skin}
    \Gamma_\mathrm{EMS}(f,\boldsymbol{\xi}|\{\overline{\boldsymbol{\xi}}_p\}_{p=1}^P) \approx \Gamma^\mathrm{max}_\mathrm{module}(f) \; G(f,\boldsymbol{\xi}|\{\overline{\boldsymbol{\xi}}_p\}_{p=1}^P)
\end{equation}
where the peak RCS of the reflector is approximated with the single SP-EMS one $\Gamma^\mathrm{max}_\mathrm{module}(f)$, while the angular and frequency selectivity is approximated by the same normalized array factor in \eqref{eq:RIS_array_factor}. Criterion (3) constraints the design of the SP-EMS reflector, that shall trade between reflectivity (large modules) and robustness to frequency selectivity (small modules).

\subsection{Design of the SP-EMS, Full-Wave Modeling, and Manufacturing Options}\label{subsect:EMS_design}

The design of SP-EMS has been widely discussed in the recent literature on the topic, with several examples already demonstrated in smart environment applications \cite{9718037,9975205,9580737}. In the following, this concept is customized to the design of retro-reflecting planar structures according to the following guidelines: \textit{(i)} the meta-atomic structure is chosen to comply with a sub-wavelength lattice (i.e., $0.3\lambda_0$ periodicity) to enable retro-reflection capability even for grazing angles ($\phi\rightarrow 90$ deg) as well as to enable proper homogenization tools to be adopted (based on the generalized sheet transition condition \cite{9718037,9975205,9580737}); \textit{(ii)} a ROGERS3003 substrate is chosen as a reference for the implementation of the meta-atom to guarantee an adequate trade-off between costs, losses, and overall efficiency in the considered bands; \textit{(iii)} an elementary meta-atomic structure based on square printed patch with variable side is implemented to demonstrate that the conceived application does not require advanced multi-layer architectures or complex patterning schemes \cite{9718037,9975205,9580737}.

Following such considerations, the design of a set of SP-EMSs has been carried out through the system-by-design approach outlined in \cite{9718037,9975205,9580737}. To this end, the meta-atom response has been firstly full-wave modeled in Ansys HFSS assuming local periodicity conditions to deduce the equivalent susceptibility tensors \cite{9718037,9975205,9580737}. The design process has been then performed to implement the a-periodic SP-EMS patterning by assuming a retro-reflection according to the generalized Snell's law \cite{9718037,9975205,9580737}. 

\begin{figure}[!t]
    \centering
    \subfloat[][SP-EMS module in HFSS]{\includegraphics[width=0.5\columnwidth]{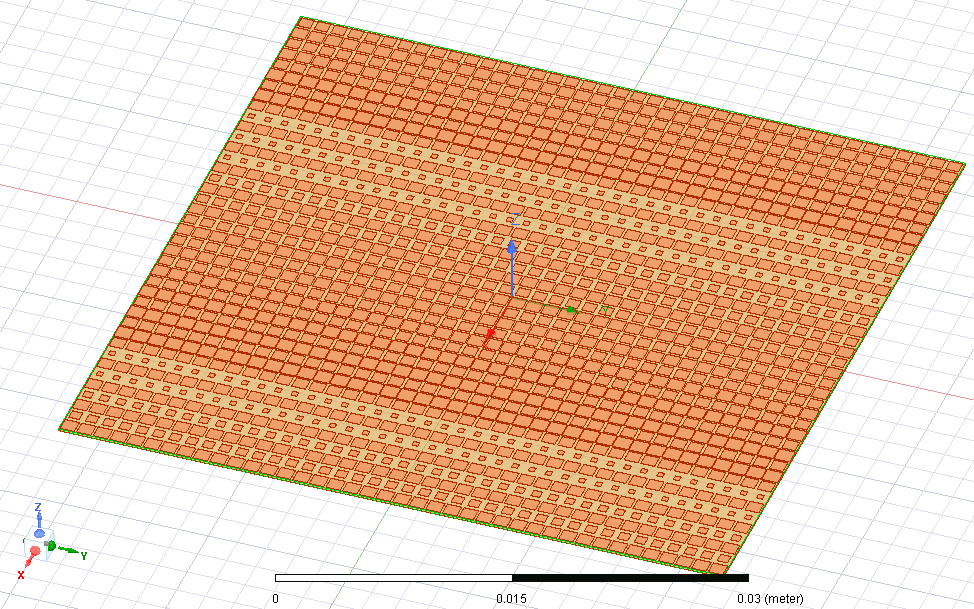}\label{subfig:SPEMS_HFSS}}\\
    \subfloat[][SP-EMS module size $4.58\times 4.58$ cm$^2$]{\includegraphics[width=0.5\columnwidth]{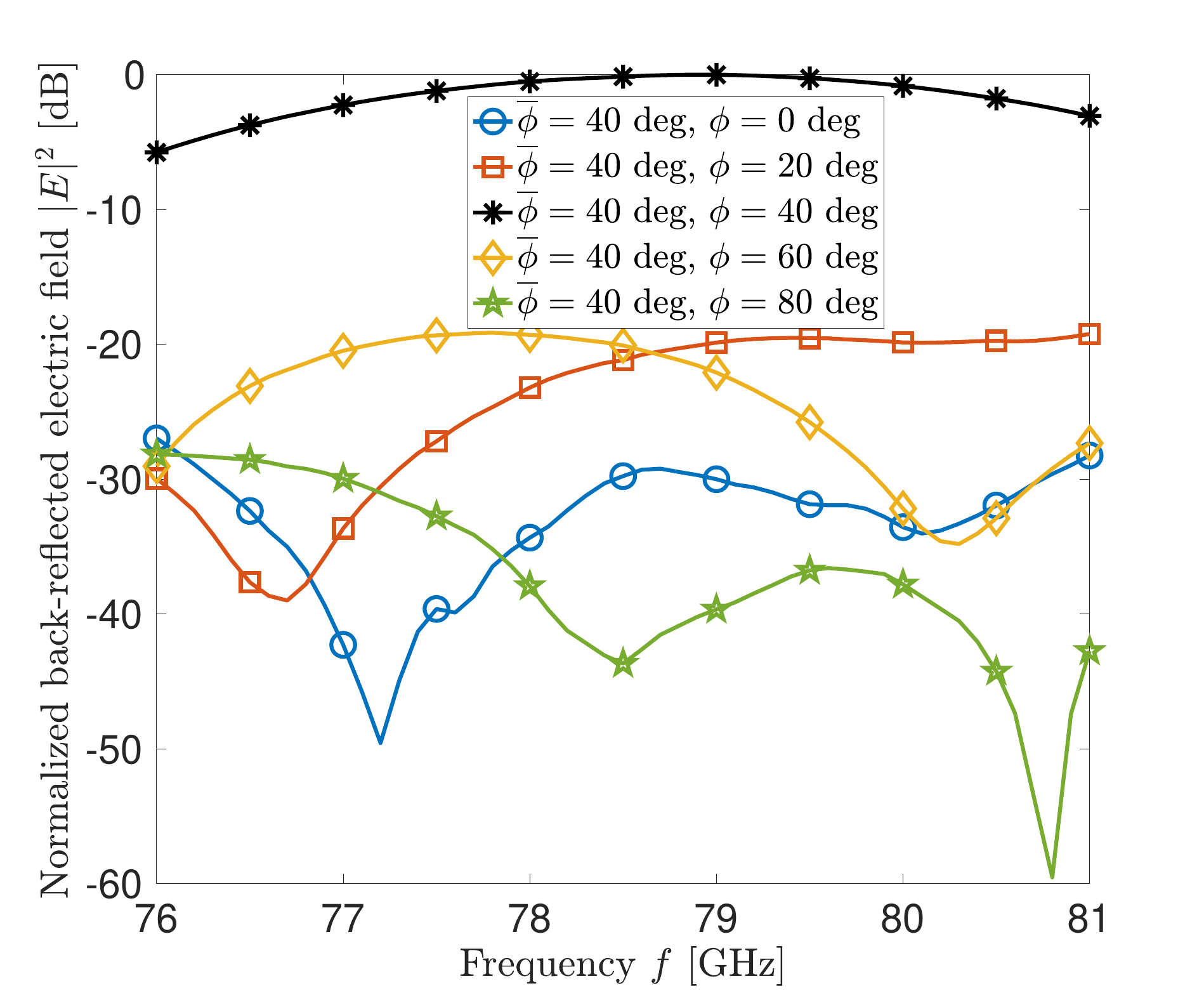}\label{subfig:Emod_5x5}}
    \subfloat[][SP-EMS module size $9.17\times 9.17$ cm$^2$]{\includegraphics[width=0.5\columnwidth]{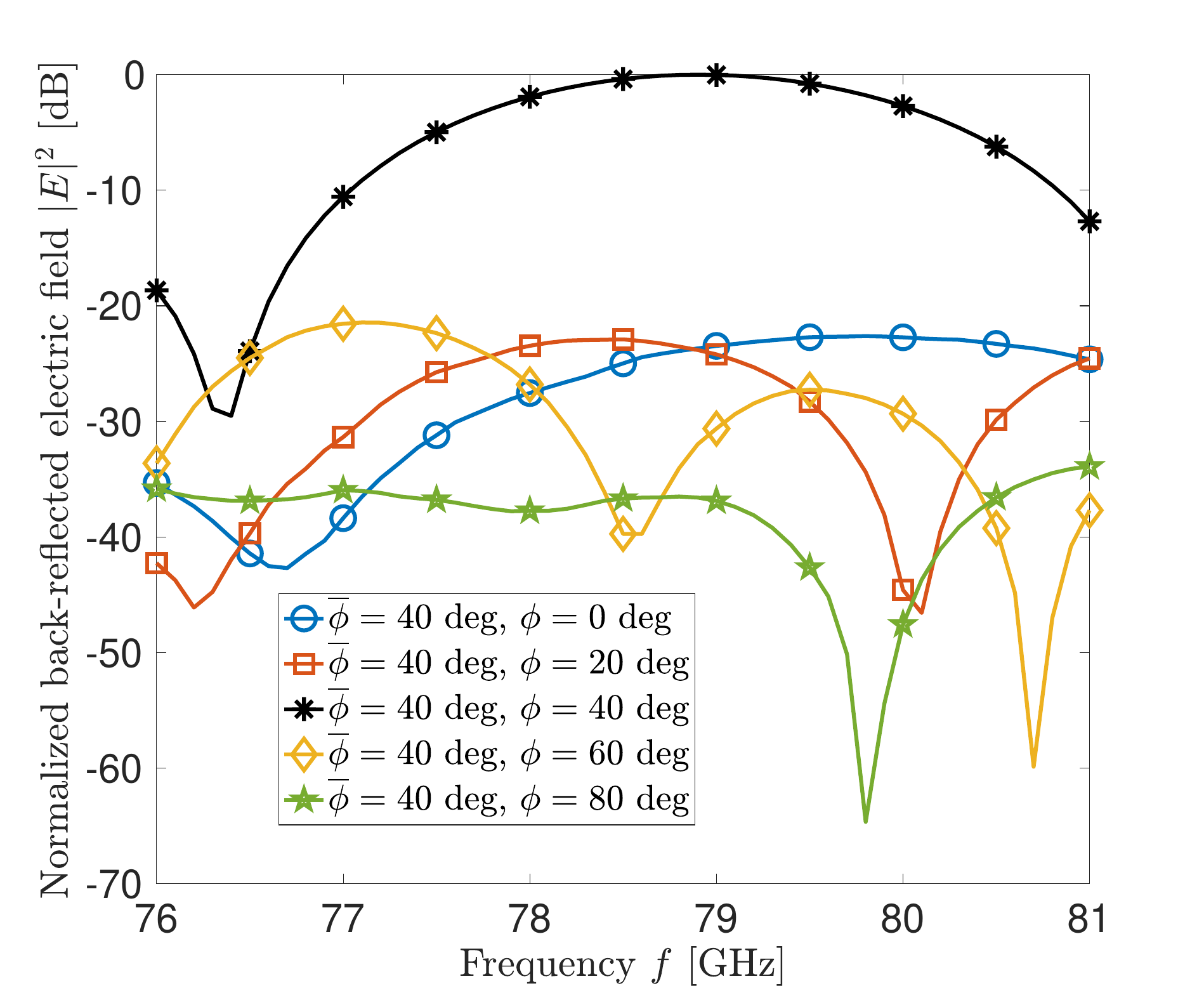}\label{subfig:Emod_10x10}} 
    \caption{HFSS model of the SP-EMS (a), normalized squared absolute value of the retro-reflected electric field, simulated with HFSS, from a (b) $4.58\times 4.58$ cm$^2$ SP-EMS and (c) $9.17\times 9.17$ cm$^2$ SP-EMS. The design process has been carried out assuming a central frequency of $f_0=78.5$ GHz, and combining $\lambda_0/3$-spaced meta atoms printed on a Rogers3003 substrate. In both cases, the SP-EMS is configured to retro-reflect the incident signal from $\overline{\boldsymbol{\xi}} = [\overline{\theta},\overline{\phi}]^T = [0, 40]^T$ deg according to the generalized Snell's law.}
    \label{fig:Emod}
\end{figure}

\begin{figure}
    \centering
    \includegraphics[width=0.7\columnwidth]{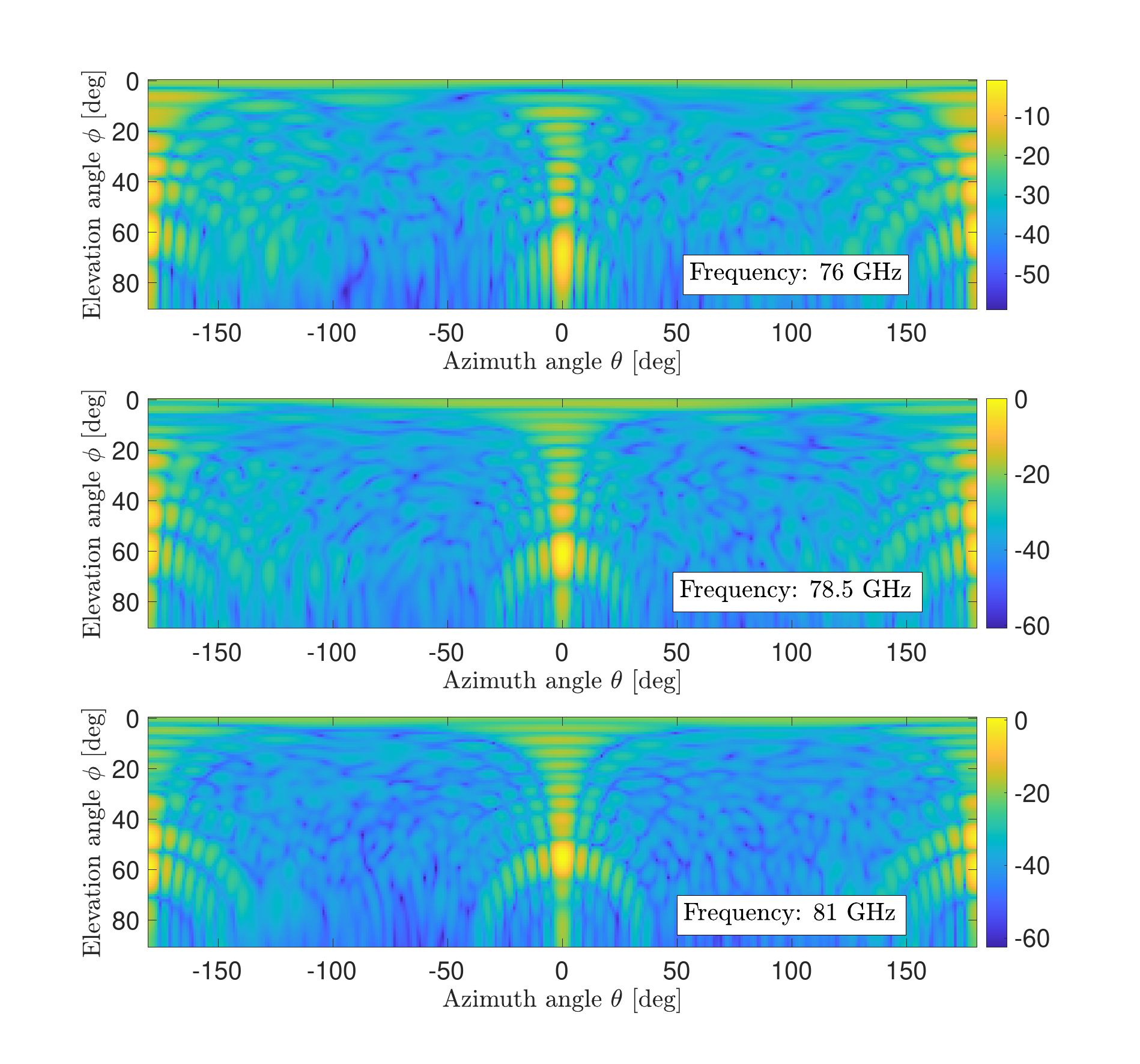}
    \caption{Normalized reflection power pattern (i.e., the squared absolute value of the reflected electric field) of a $4.58\times 4.58$ cm$^2$ SP-EMS when illuminated from a fixed direction $\boldsymbol{\xi} = [\theta,\phi]^T = [0, 60]^T$ deg and phase-configured to retro-reflect at angles $\overline{\boldsymbol{\xi}} = [\overline{\theta},\overline{\phi}]^T = [0, 60]^T$ deg. }
    \label{fig:pattern}
\end{figure}

The full-wave results of the designed finite SP-EMSs obtained using Ansys HFSS are shown in Figs \ref{fig:Emod} and \ref{fig:pattern}. In such a case, a plane wave illumination has been enforced, and the patterned structure has been modeled assuming a finite element-boundary integral (FE-BI) formulation to account for the edge effects and diffraction properties. In Fig. \ref{fig:Emod}, we show the squared absolute value of the retro-reflected electric field (normalized to its maximum value) as a function of the baseband frequency $f\in[76,81]$ GHz, for a module of $4.58\times 4.58$ cm$^2$ size (Fig. \ref{subfig:Emod_5x5}) and of $9.17 \times 9.17$ cm$^2$ (Fig. \ref{subfig:Emod_10x10}), both configured to retro-reflect at $\overline{\boldsymbol{\xi}} = [\overline{\theta},\overline{\phi}]^T = [0, 40]^T$ deg. The squared modulus of the reflected electric field is proportional to the monostatic RCS of the module \cite{skolnik}, and can be used to assess the validity of the analytical model \eqref{eq:RCS_skin} in predicting the frequency-selectivity of the EMSs, either SP-EMSs or RISs.  
Comparing black curves in \ref{fig:Emod}, showing the module spectral response for $\phi=\overline{\phi}=40$ deg, and the related curves in Fig. \ref{fig:RCS_vs_frequency}, we notice a difference of few dB in both cases. Therefore, the analytical results predicted by \eqref{eq:RIS_array_factor} show good adherence to the HFSS design in terms of frequency selectivity pattern (i.e., the trend of the RCS varying the baseband frequency $f$) and can be used for a proper performance evaluation of the localization performance, as discussed in the next Section \ref{sect:results}. Fig. \ref{fig:pattern}, instead, shows the power reflection pattern, proportional to the bistatic RCS, of a $4.58\times 4.58$ cm$^2$ SP-EMS module illuminated from $\boldsymbol{\xi} = [\theta,\phi]^T=[0,60]$ deg. As expected, the elevation angle of maximum reflection (i.e., the angular index of maximum reflected field in Fig. \ref{fig:pattern}) shifts from $\phi_{\mathrm{peak}} = 70$ deg ($f=76$ GHz) to $\phi_{\mathrm{peak}} = 56$ deg ($f=81$ GHz), while the azimuth does not change, as can be demonstrated. A similar trend is observed for the reflection beamwidth. For $f=76$ GHz, the reflection beamwidth along elevation is $\Delta \phi = 20$ deg, while at $f=81$ GHz it narrows to $\Delta \phi = 11$ deg (at $f_0=78.5$ GHz it amounts to $\Delta \phi = 15$ deg). In all cases, we have $\Delta \theta = 10$ deg. For the selected module size, the angle of peak reflectivity $\phi_{peak}$ falls within the -3 dB reflection beamwidth $\Delta \phi$ $\forall f\in[76,81]$ GHz, guaranteeing that the SP-EMS module is capable of retro-reflecting the signal without significant frequency-dependent power loss. With single modules of $4.58\times 4.58$ cm$^2$ size and center bandwidth reflection beamwidth of $\Delta \theta = 10$ deg, $\Delta \phi = 15$ deg, we roughly need $36 \times 6 = 216$ modules to implement an omnidirectional retro-reflector, thus a composite SP-EMS of $60$ cm side. Differently, employing modules of a larger size to increase its RCS, e.g., $9.17 \times 9.17$ cm$^2$, implies having a much larger footprint (the single module is larger and it also provides a narrower reflection beamwidth).

\section{Localization Performance}\label{sect:results}

\begin{figure*}[!t]
\centering
    \subfloat[][$3\sigma = 0.5$ m, $B=1$ GHz]{\includegraphics[width=0.5\columnwidth]{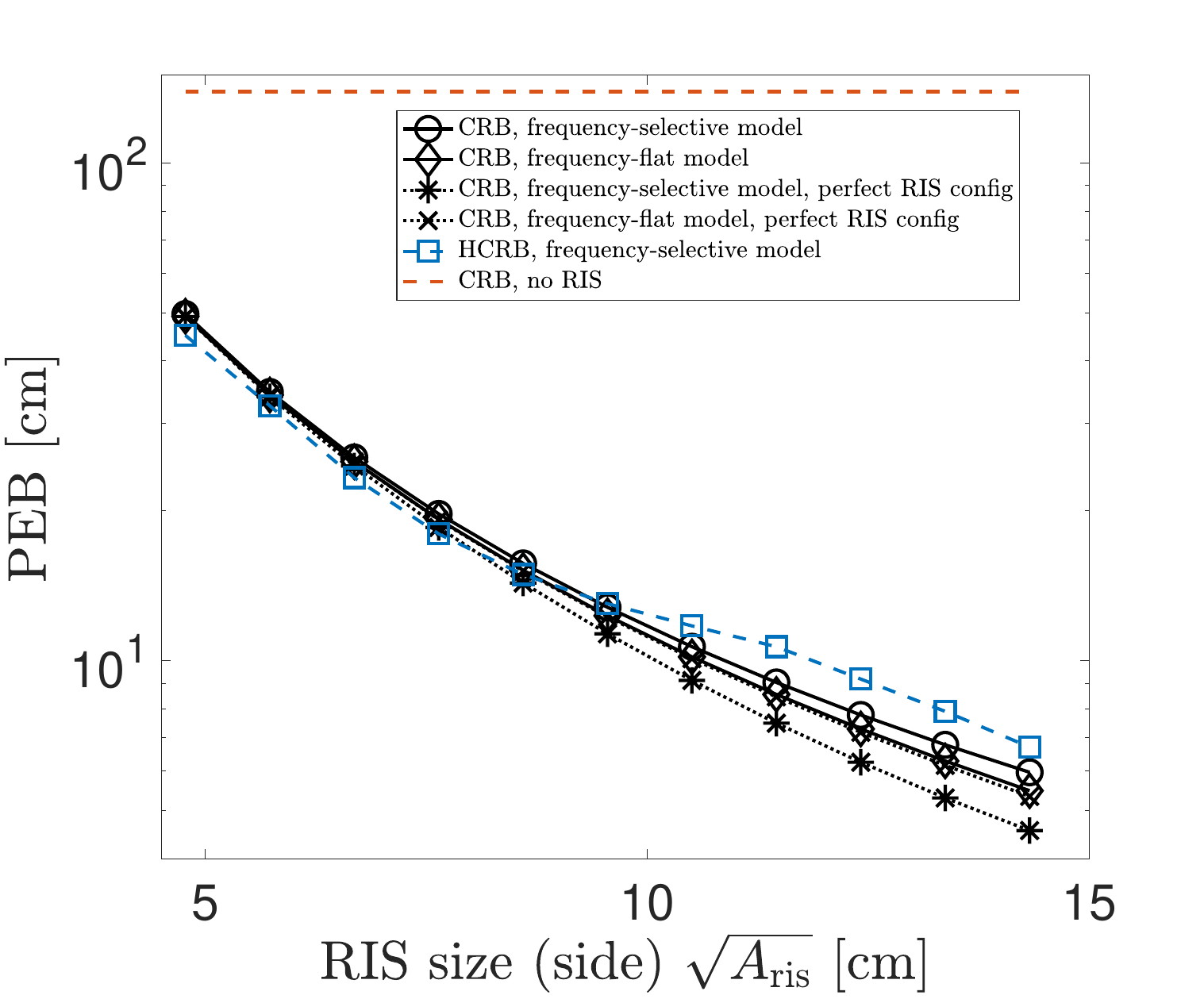}\label{subfig:CRB_0dot5_B1G}} 
    \subfloat[][$3\sigma = 0.5$ m, $B=4$ GHz]{\includegraphics[width=0.5\columnwidth]{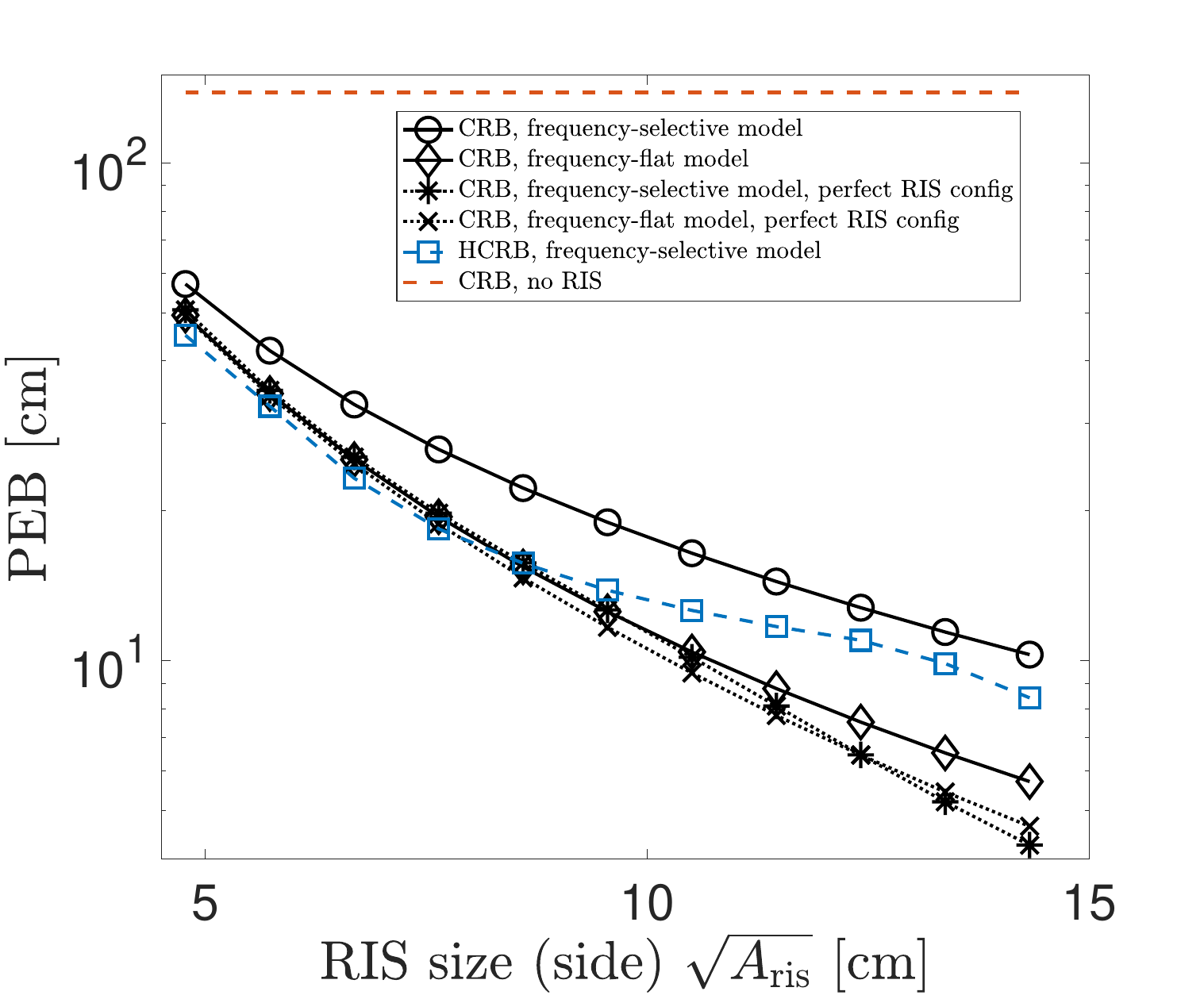}\label{subfig:CRB_0dot5_B4G}} \\
    \subfloat[][$3\sigma = 2$ m, $B=1$ GHz]{\includegraphics[width=0.5\columnwidth]{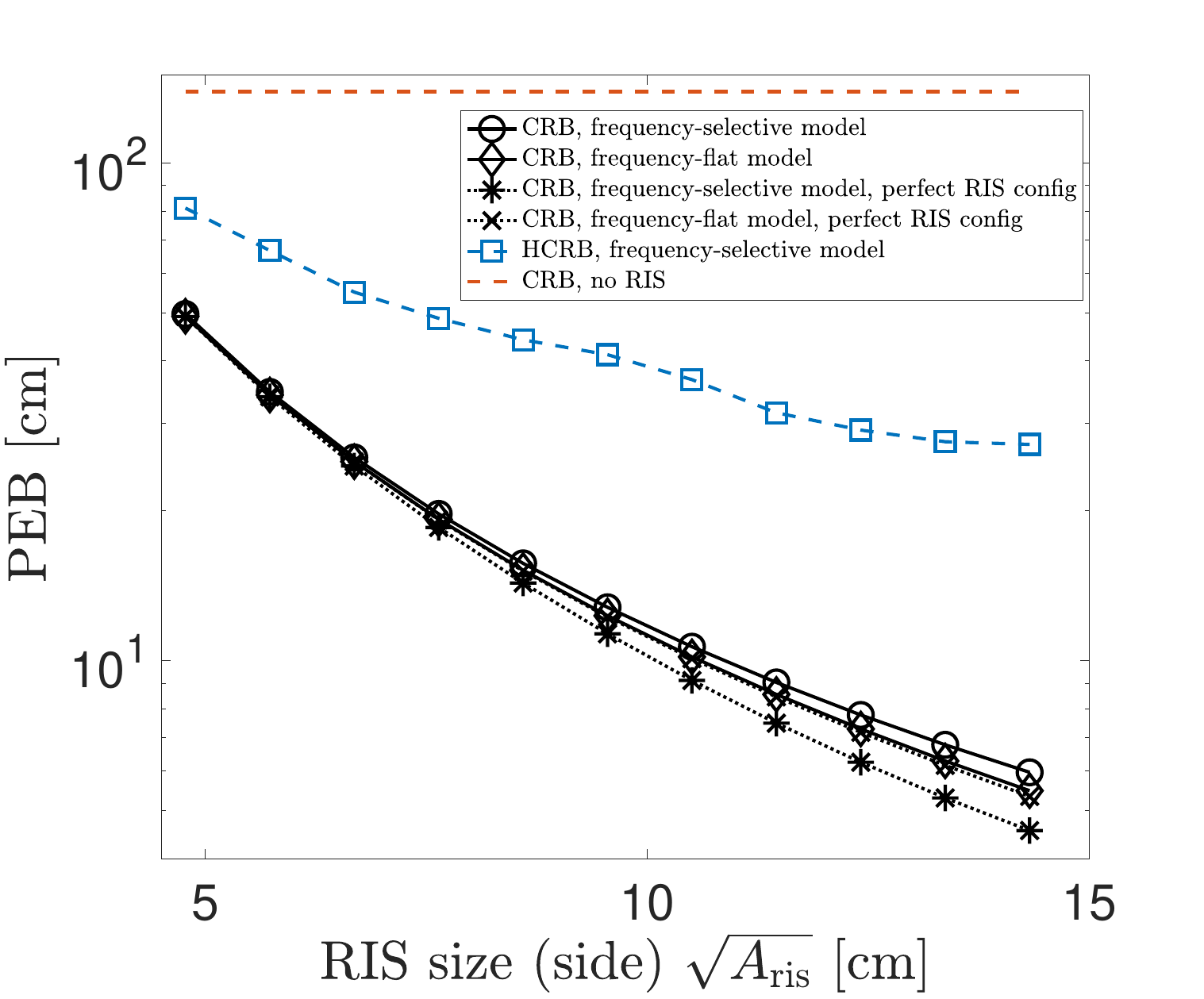}\label{subfig:CRB_2_B1G}} 
    \subfloat[][$3\sigma = 2$ m, $B=4$ GHz]{\includegraphics[width=0.5\columnwidth]{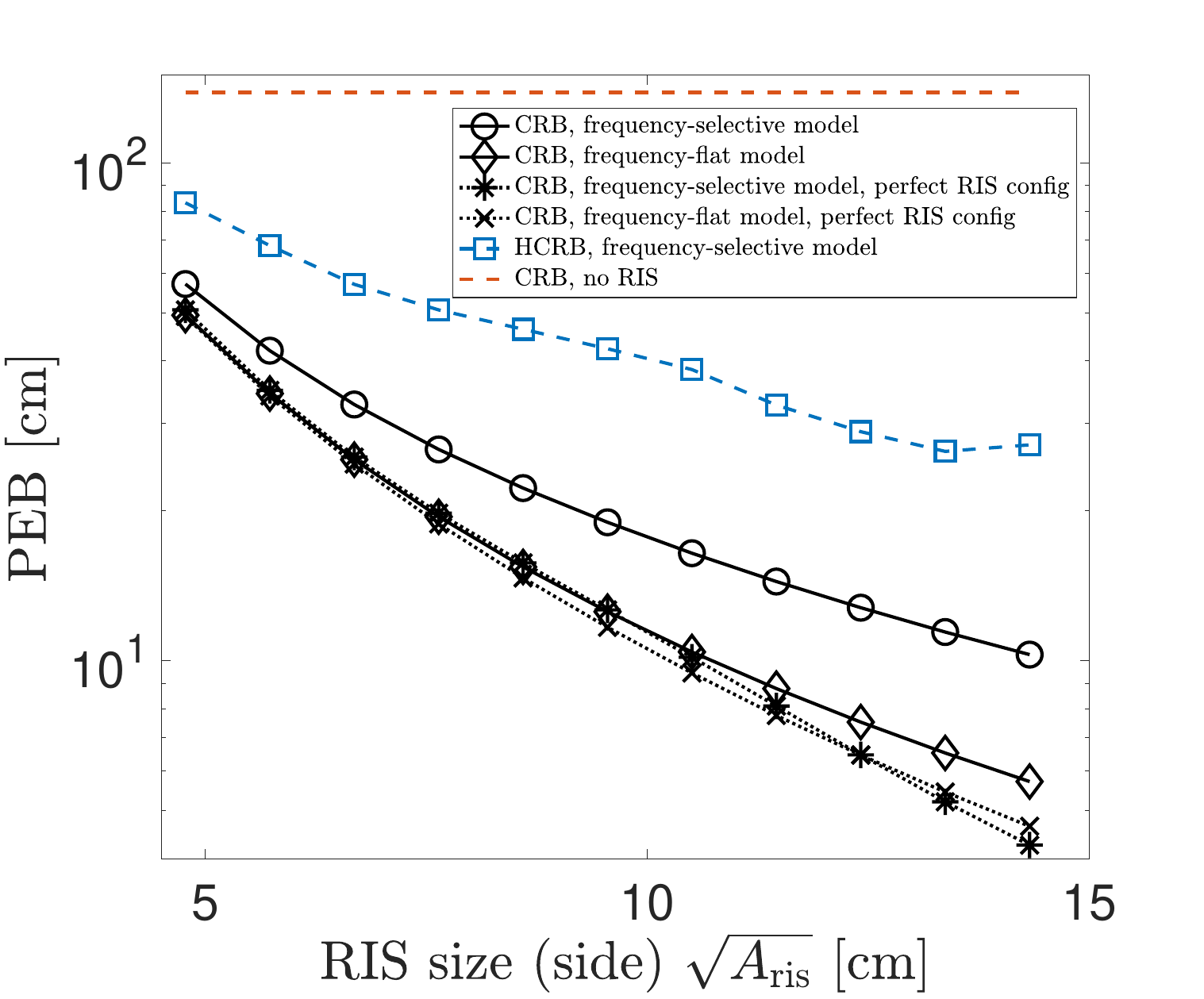}\label{subfig:CRB_2_B4G}}
    \caption{PEB vs. RIS size for (a) $3\sigma = 0.5$ m, $B=1$ GHz, (b) $3\sigma = 0.5$ m, $B=4$ GHz, (c) $3\sigma = 2$ m, $B=1$ GHz, (d) $3\sigma = 2$ m, $B=4$ GHz. }
    \label{fig:HCRB_CRB_vs_M}
\end{figure*}

\begin{figure}
    \centering
    \includegraphics[width=0.5\columnwidth]{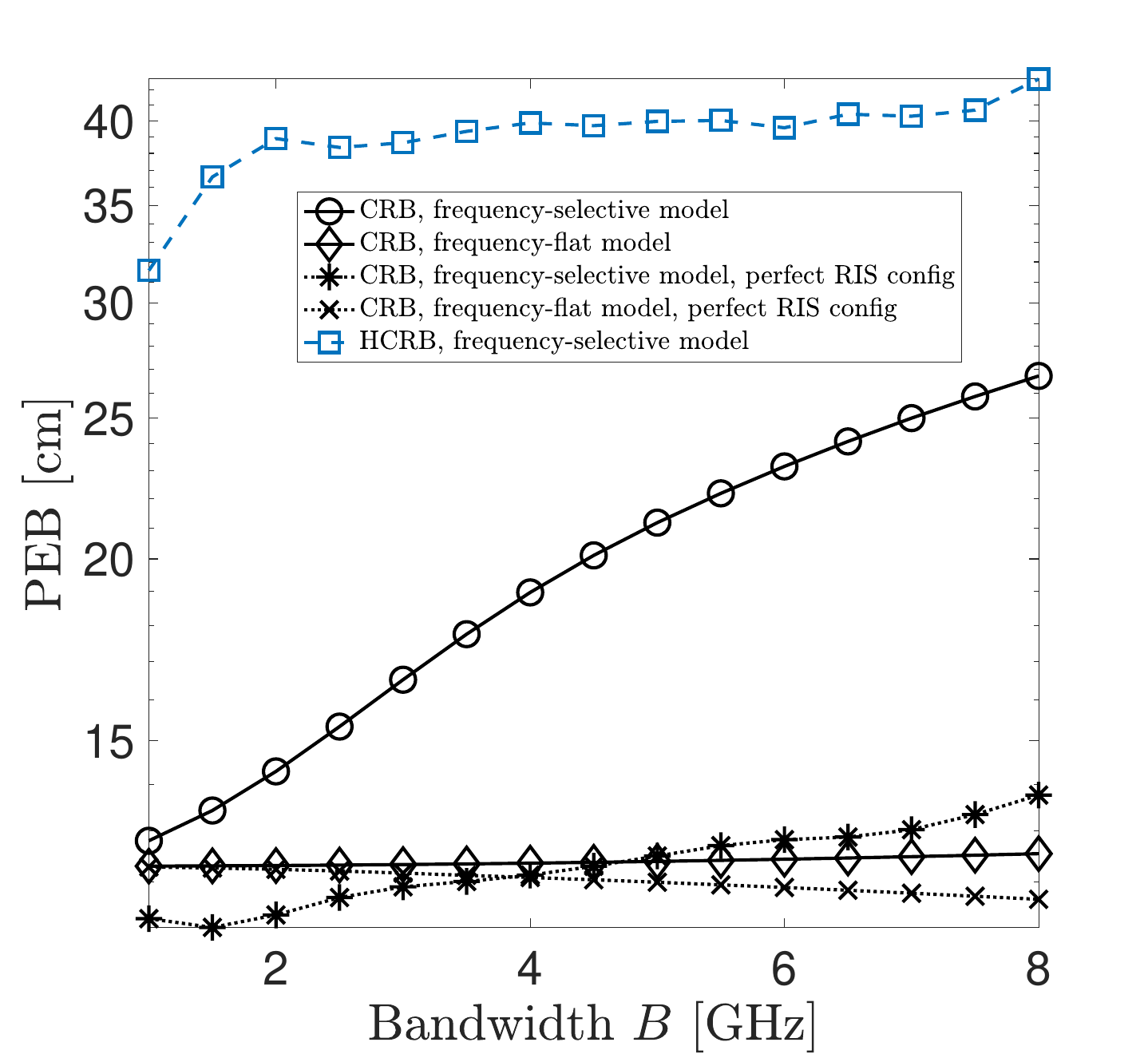}
    \caption{PEB vs. bandwidth $B$.}
    \label{fig:HCRB_CRB_vs_B}
\end{figure}

This section shows numerical results quantifying the localization performance bounds when the vehicle is equipped with a RIS. The same results would apply to the SP-EMS case; we refer to RIS for simplicity. To this aim, we evaluate the CRB on carrier-phase-based RIS's position estimation in the following cases: \textit{(i)} perfect RIS configuration, i.e., the configuration angles $\overline{\boldsymbol{\xi}}$ are known and shall not be estimated and \textit{(ii)} imperfect RIS configuration, i.e., the configuration angles $\overline{\boldsymbol{\xi}}$ are part of the parameters to be estimated. The CRB is evaluated in both wideband and narrowband modeling assumptions, whereas the latter is reported as benchmark and assumes 
\begin{equation}\label{eq:narrowband_assumption}
    s\left(t \hspace{-0.1cm}- \hspace{-0.1cm}2 \tau_0\hspace{-0.1cm} - \hspace{-0.1cm}\Delta\tau^i_{\ell}\hspace{-0.1cm} - \hspace{-0.1cm}\Delta\tau^o_{\ell} \hspace{-0.1cm}- \hspace{-0.1cm}2 \Delta\tau_{nm}\right) \approx s\left(t\hspace{-0.1cm}- \hspace{-0.1cm}2 \tau_0\right)
\end{equation}
in \eqref{eq:Rx_signal}. Furthermore, to model the a-priori knowledge on the RIS position, that affects its configuration angles, we compute the HCRB for deterministic position $\mathbf{x}$ and random configuration angles $\overline{\boldsymbol{\xi}}$, characterized by distribution $p(\overline{\boldsymbol{\xi}})=p(\widehat{\boldsymbol{\xi}})$. Details on CRB and HCRB derivations are reported in Appendix \ref{Appendix:HCRB}. The employed metric is the position error bound (PEB), defined as
\begin{equation}
    \mathrm{PEB} = \sqrt{\frac{\mathrm{trace\left([\mathbf{C}]_{1:3,1:3}\right)}}{3}} 
\end{equation}
where $\mathbf{C}$ is either $\mathbf{C}_\mathrm{HCRB}$ (for HCRB), $\mathbf{C}_\mathrm{CRB}$ (for CRB) or $\mathbf{C}^\mathrm{u}_\mathrm{CRB}$ (for CRB and perfect RIS configuration).
As a performance benchmark, we also show the PEB achieved without the usage of a dedicated EMS (RIS or SP-EMS). This PEB is obtained with the empirical criterion described by \eqref{eq:RCS_RIS_criterion} (Section \ref{sect:RIS_reflector_design}). We consider that the experienced RCS of the bare vehicle is at least 10 dB less w.r.t. the one of a RIS of $7.5 \times 7.5$ cm$^2$ size. Moreover, we take into account the fact that the strongest back-scattering point on the vehicle does not coincide with the desired location of the RIS on the roof, due to perspective deformations. This latter effect has been simulated with RemCom Wavefarer software \cite{wavefarer} and acts a \textit{bias} on the position estimate. 

Concerning simulation parameters, we consider the sensing terminal operating at $f_0=78.5$ GHz, on a variable bandwidth $B\in[1,8]$ GHz,  equipped with $1$ Tx antenna and $20\times 20$ Rx antennas along $y$ and $z$ axes. The RIS is square and made by $N=M=\sqrt{A_\mathrm{ris}/d^2}\in[50,150]$ elements at $\lambda_0/4$, corresponding to a size ranging from $4.7$ cm to $14.3$ cm. For the evaluation of the HCRB, we assume an a-priori knowledge of the configuration angles $\overline{\boldsymbol{\xi}}$ coming from a GPS-aided position estimation with uncertainty $3\sigma = 0.5$ m (as for an accurate RTK setup) and $3\sigma = 2$ m, as for a off-the-shelf GPS without urban canyoning effects. The RIS position $\mathbf{x}$ is instead treated as a deterministic parameter\footnote{The RIS position $\mathbf{x}$ can also be treated as a random parameter with a Gaussian a-priori distribution $p(\widehat{\mathbf{x}})$, computing the Bayesian CRB (BCRB). Both HCRB and BCRB are viable options.}. In all the evaluations, we have the sensing terminal in the origin of the reference system and the RIS in $\mathbf{x} = [10,5,-6.5]^T$ m, that is a random position within the sensing coverage area. The sensing terminal emits $23$ dBm of average power (for each measurement channel) and the Rx signal is corrupted by thermal noise, $N_0 = -173$ dBm/Hz. 

\subsection{Results}

The first set of results is summarized in Fig. \ref{fig:HCRB_CRB_vs_M}. Here, we report both the HCRB and the CRB performance varying the RIS side $\sqrt{A_\mathrm{ris}}$. For CRB, we also show the performance of the narrowband model, namely assuming $\Gamma_\mathrm{ris}(f_0, \boldsymbol{\xi}|\overline{ \boldsymbol{\xi}})$, no frequency selectivity of the reflection coefficient $\beta$. This is the commonly assumed framework in most of the available literature (e.g., see \cite{Buzzi_RISforradar_journal,9625826}). Let us focus on the case $3\sigma = 0.5$ m, depicted in Figs. \ref{subfig:CRB_0dot5_B1G} and \ref{subfig:CRB_0dot5_B4G} for $B=1$ GHz and $B=4$ GHz, respectively. In the first case, the PEB provided by the CRB changes its behavior according to the knowledge (or not) of the configuration angles of the RIS $\overline{\boldsymbol{\xi}}$. For unknown (but deterministic) $\overline{\boldsymbol{\xi}}$ (solid black lines), the PEB of the narrowband model is lower than the PEB of the wideband one. This is expected, as no wideband modeling implies no filtering effects on the Tx signal due to the RIS (see Section \ref{subsect:posErr}) and therefore a higher Rx power. Differently, for $\overline{\boldsymbol{\xi}}=\boldsymbol{\xi}$ (perfect knowledge, dotted black lines), the PEB of the narrowband model is \textit{higher} than the PEB of the wideband one. In this latter case, when the phase of the RIS does not need to be estimated, (e.g., after the RIS alignment procedure as detailed in Section \ref{subsect:posErr}), the \textit{position-dependent} frequency response of the sensing channel (via $\Gamma_\mathrm{ris}(f, \boldsymbol{\xi}|\overline{ \boldsymbol{\xi}})$) introduces a non-negligible amount of information that allows compensating for the loss of Rx power due to the low-pass nature of $\Gamma_\mathrm{ris}(f, \boldsymbol{\xi}|\overline{ \boldsymbol{\xi}})$. This effect is peculiar of a sensing system observing a RIS-equipped target, and can be clearly observed for $B=1$ GHz (Fig. \ref{subfig:CRB_0dot5_B1G}) and also for $B=4$ GHz (Fig. \ref{subfig:CRB_0dot5_B4G}), where, however, the filtering effect is stronger. Differently, the PEB provided by the HCRB (dashed blue curves) can be tighter than the PEB provided by CRB when the a-priori information is significant, as for $3\sigma = 0.5$ m. For higher inaccuracy on the angle estimation ($3\sigma = 2$ m, Figs. \ref{subfig:CRB_2_B1G} and \ref{subfig:CRB_2_B1G}) only the HCRB is affected, as expected. In general, HCRB provides a more conservative bound that better represents the case where one or more parameters are inherently random, as for $\overline{\boldsymbol{\xi}}$. In any case, we notice that the usage of a EMS consistently improves the localization performance compared to the bare vehicle case. The red dashed curve in Fig. \ref{fig:HCRB_CRB_vs_M}, representing the performance we can expect without the RIS/SP-EMS, always exhibits the highest PEB.

The trend of the PEB varying the employed bandwidth $B$ is instead reported in Fig. \ref{fig:HCRB_CRB_vs_B}, for $N=M=\sqrt{A_\mathrm{ris}/d^2}=100$ (thus a $9.5\times 9.5$ cm$^2$ RIS) and $3\sigma=2$ m. The PEB by CRB for frequency-dependent RCS $\Gamma_\mathrm{ris}(f, \boldsymbol{\xi}|\overline{ \boldsymbol{\xi}})$ monotonically increases with $B$. The former effect is the immediate consequence of the low-pass filtering of $\Gamma_\mathrm{ris}(f, \boldsymbol{\xi}|\overline{ \boldsymbol{\xi}})$ on the incident signal, that reduces the retro-reflected energy in the direction of the sensing terminal. Differently, for narrowband system modeling, the PEB does not appreciably decrease with $B$. This effect is  typical of carrier-phase positioning systems (e.g., see \cite{wymeersch2023fundamental}), and it is due to the effective bandwidth of the Tx signal $s(t)$, that is \cite{BigS}:
\begin{equation}\label{eq:eff_band}
    B_\mathrm{eff}^2 = f_0^2 + \frac{B^2}{12} \approx f_0^2,
\end{equation}
largely dominated by the carrier component for the considered settings, and almost insensitive to $B$ (for $B=10$ GHz and $f_0=77$ GHz, the first term of \eqref{eq:eff_band} is 3 orders of magnitude higher than the second). For $\overline{\boldsymbol{\xi}}=\boldsymbol{\xi}$ (perfect RIS configuration), instead, the wideband model again provides an increasing PEB with $B$, that however, for $B\leq 4$ GHz, outperforms the narrowband model. This is the effect observed in Fig. \ref{fig:HCRB_CRB_vs_M}. For $B > 4 $ GHz, however, the information brought to the sensing system by the position-dependent frequency pattern of $\Gamma_\mathrm{ris}(f, \boldsymbol{\xi}|\overline{\boldsymbol{\xi}})$ is not sufficient to compensate for the severe power loss and the trend swaps. The HCRB, instead, increases for $B\leq 2$ GHz, but for larger values the bandwidth $B$ plays no significant role, as the PEB is approximately dominated by the uncertainty on $\overline{ \boldsymbol{\xi}}$. 

\section{Conclusion}\label{sect:conclusion}

This paper proposes lodging EMSs on vehicle roofs to increase their radar visibility and to make extended targets behave as a point targets, easing their detection and tracking from sensing data. EMSs act as EM markers for impinging sensing signals, enabling intentional retro-reflections towards the sensing terminal (e.g., ISAC BS or MIMO radar). We detail the design on a RIS-based reflector considering the spatially wideband effect, that produces an in-band filtering on the impinging sensing signal, and the imperfect RIS phase configuration, that causes a drastic reduction of the experienced RIS's RCS. We propose an adaptive RIS alignment procedure to configure the RIS phase for retro-reflection, that caters with mobility. Furthermore, to reduce the implementation cost associated to a RIS-based reflector, we also outline a possible cost-effective realization based on multiple pre-configured SP-EMSs modules, whose design is validated with full-wave simulations carried out in Ansys HFSS. Finally, localization performance of a vehicle equipping a retro-reflecting EMS is assessed in terms of CRB and HCRB in both spatially wideband and narrowband modeling assumptions, the latter used as benchmark, showing the benefits brought by the employment of a retro-reflective EMS w.r.t. to the case of a bare vehicle.  

\appendices

\section{Hybrid CRB Calculation}\label{Appendix:HCRB}

For the evaluation of the HCRB, we make reference to the wideband, frequency-dependent signal model in \eqref{eq:Rx_signal_frequency}. The observations from the $L$ channels can be stacked into a vector, obtaining 
\begin{equation}
    \mathbf{y}(f) = \mathbf{a}(f,\boldsymbol{\theta})  + \mathbf{z}(f)\sim\mathcal{CN}(\mathbf{a}(f,\boldsymbol{\theta}), N_0\mathbf{I}_L\delta(f))
\end{equation}
where $\mathbf{a}(f,\boldsymbol{\theta})\in\mathbb{C}^{L\times 1}$ is the non-linear model relating the parameters to be estimated 
\begin{equation}\label{eq:parameters}
\boldsymbol{\theta}=[
    \mathbf{x}^T,\;
    \overline{\boldsymbol{\xi}}^T]^T\in\mathbb{R}^{6\times 1}
\end{equation}
to the observation $\mathbf{y}(f)$, and $\mathbf{z}(f)$ is the additive noise. The set of parameters $\boldsymbol{\theta}$ is herein split between deterministic, namely the RIS position $\mathbf{x}$, and random, namely the RIS angular configuration $\overline{\boldsymbol{\xi}}$. We make the assumption that the orientation of the RIS in space, i.e., angle $\psi$, is known and does not need to be estimated. The same applies to the amplitude and phase of the scattering coefficient $\rho$ in \eqref{eq:Rx_signal}. These assumptions allows retrieving an optimistic PEB. However, the following derivation is general and can be extended to an unknown (but deterministic) orientation of the RIS and scattering amplitude. Notice that modelling $\overline{\boldsymbol{\xi}}$ as a random parameter to be estimated, with an a-priori knowledge of its distribution, is only one of the choices, following from the initial estimation of the RIS position at the vehicle. Alternatively, $\overline{\boldsymbol{\xi}}$ can be assumed as deterministic, whose true value $\overline{\boldsymbol{\xi}}=\boldsymbol{\xi}$ corresponding to the optimal RIS configuration is unknown. 

The hybrid information matrix (HIM) $\mathbf{J}_{\mathrm{HCRB}}$ is computed as follows:
\begin{equation}\label{eq:HCRB_FIM}
    \mathbf{J}_{\mathrm{HCRB}} = \mathbf{J}_{\mathrm{D}} = \mathbb{E}_{\overline{\boldsymbol{\xi}}} \left[ \mathbf{F}\right] + \mathbf{J}_{\mathrm{R}}
\end{equation}
where the first term is the deterministic component of the HIM, obtained by taking the expectation of the Fisher information matrix (FIM) $\mathbf{F}$ over the random parameters $\overline{\boldsymbol{\xi}}$ while $\mathbf{J}_{\mathrm{R}}$  is the additional HIM component due to a-priori knowledge on the random parameters $\overline{\boldsymbol{\xi}}$. The computation of $\mathbf{J}_{\mathrm{R}}$ requires $p(\overline{\boldsymbol{\xi}})=p(\widehat{\boldsymbol{\xi}})$, i.e., the PDF of the estimated incidence angles onto the RIS, obtained from the position estimation in Section \ref{sect:RIS_reflector_design} knowing the heading $\psi$.

FIM $\mathbf{F}$ is block-partitioned as follows
\begin{equation}\label{eq:FIM_HCRB}
    \mathbf{F} = \begin{bmatrix}
        \mathbf{F}_{\mathbf{x}\mathbf{x}}& 
         \mathbf{F}_{\mathbf{x}\overline{\boldsymbol{\xi}}}\\
        \mathbf{F}_{\overline{\boldsymbol{\xi}}\mathbf{x}}&  \mathbf{F}_{\overline{\boldsymbol{\xi}}\overline{\boldsymbol{\xi}}}
    \end{bmatrix}
\end{equation}
with straightforward dimensions, whose entries are
\begin{align}
    \mathbf{F}_{\boldsymbol{\mu}\boldsymbol{\nu}} =  \frac{2}{N_0} \Re \left\{ \int\limits_{-B/2}^{B/2} \left(\frac{\partial \mathbf{a}(f,\boldsymbol{\theta})}{\partial \boldsymbol{\mu}}\right)^H \frac{\partial \mathbf{a}(f,\boldsymbol{\theta})}{\partial  \boldsymbol{\nu}} df \right\}
\end{align}
where $\boldsymbol{\mu}$ and $\boldsymbol{\nu}$ can be any set of the parameters to be estimated, i.e., $\mathbf{x}$, $\overline{\boldsymbol{\xi}}$. Single FIM terms are:
    \begin{equation}\label{eq:partial_a_partial_x}
        \begin{split}
             \frac{\partial \mathbf{a}(f,\boldsymbol{\theta})}{\partial \mathbf{x}}  =  S(f) & \left[  \frac{\partial \beta(f,\mathbf{x},\gamma,\overline{\boldsymbol{\xi}})}{\partial \mathbf{x}} e^{-j 2 \pi (f_0+f) (2\tau_0 + \Delta\boldsymbol{\tau}_i + \Delta\boldsymbol{\tau}_o)} + \right.\\
             & \left.\beta(f,\mathbf{x},\gamma,\overline{\boldsymbol{\xi}}) \frac{\partial}{\partial \mathbf{x}} \left(e^{-j 2 \pi (f_0+f) (2\tau_0 +\Delta\boldsymbol{\tau}_i + \Delta\boldsymbol{\tau}_o)}\right)\right] \in\mathbb{C}^{L\times 3}
        \end{split}
    \end{equation}
\begin{equation} \label{eq:partial_a_partial_xi}       
            \frac{\partial \mathbf{a}(f,\boldsymbol{\theta})}{\partial \overline{\boldsymbol{\xi}}} =  S(f)  \frac{\partial \beta(f,\mathbf{x},\gamma,\overline{\boldsymbol{\xi}})}{\partial \overline{\boldsymbol{\xi}}} e^{-j 2 \pi (f_0+f) (2\tau_0 +\Delta\boldsymbol{\tau}_i+\Delta\boldsymbol{\tau}_o)}  \in\mathbb{C}^{L\times 2}.
\end{equation}%
The inverse of the FIM $\mathbf{F}$ provides the CRB for deterministic parameters, namely $\mathbf{C}_\mathrm{CRB} = \mathbf{F}^{-1}$.


\subsection{CRB for perfect RIS configuration $\overline{\boldsymbol{\xi}}=\boldsymbol{\xi}$}
A further useful performance bound is to consider the perfect configuration of the RIS ($\overline{\boldsymbol{\xi}}=\boldsymbol{\xi}$), e.g., achieved with a reflection beam sweeping procedure in Section \ref{sect:RIS_reflector_design}. In this case, the only parameter to be estimated is $\mathbf{x}$, as the residual uncertainty on $\overline{\boldsymbol{\xi}}$ is zero, or it can be approximated as zero. The HCRB therefore degenerates to the CRB for the FIM:
\begin{equation}\label{eq:FIM_CRB}
    \mathbf{F}^{\mathrm{u}}= 
        \mathbf{F}_{\mathbf{x}\mathbf{x}}
\end{equation}
and we have $\mathbf{C}_{\mathrm{HCRB}} \succeq \mathbf{C}^\mathrm{u}_{\mathrm{CRB}} =  (\mathbf{F}^{\mathrm{u}})^{-1}$ as the uncertainty of the RIS configuration angles is removed from the FIM in \eqref{eq:FIM_HCRB} and cannot affect the estimation of the other parameters.


\bibliographystyle{IEEEtran}
\bibliography{Bibliography}

\begin{thebibliography}{10}
\providecommand{\url}[1]{#1}
\csname url@samestyle\endcsname
\providecommand{\newblock}{\relax}
\providecommand{\bibinfo}[2]{#2}
\providecommand{\BIBentrySTDinterwordspacing}{\spaceskip=0pt\relax}
\providecommand{\BIBentryALTinterwordstretchfactor}{4}
\providecommand{\BIBentryALTinterwordspacing}{\spaceskip=\fontdimen2\font plus
\BIBentryALTinterwordstretchfactor\fontdimen3\font minus
  \fontdimen4\font\relax}
\providecommand{\BIBforeignlanguage}[2]{{%
\expandafter\ifx\csname l@#1\endcsname\relax
\typeout{** WARNING: IEEEtran.bst: No hyphenation pattern has been}%
\typeout{** loaded for the language `#1'. Using the pattern for}%
\typeout{** the default language instead.}%
\else
\language=\csname l@#1\endcsname
\fi
#2}}
\providecommand{\BIBdecl}{\relax}
\BIBdecl

\bibitem{9349624}
W.~Jiang, B.~Han, M.~A. Habibi, and H.~D. Schotten, ``The road towards 6g: A
  comprehensive survey,'' \emph{IEEE Open Journal of the Communications
  Society}, vol.~2, pp. 334--366, 2021.

\bibitem{9815783}
H.~Wymeersch, A.~Pärssinen, T.~E. Abrudan, A.~Wolfgang, K.~Haneda,
  M.~Sarajlic, M.~E. Leinonen, M.~F. Keskin, H.~Chen, S.~Lindberg, P.~Kyösti,
  T.~Svensson, and X.~Yang, ``6g radio requirements to support integrated
  communication, localization, and sensing,'' in \emph{2022 Joint European
  Conference on Networks and Communications \& 6G Summit (EuCNC/6G Summit)},
  2022, pp. 463--469.

\bibitem{9625032}
M.~A. Uusitalo, P.~Rugeland, M.~R. Boldi, E.~C. Strinati, P.~Demestichas,
  M.~Ericson, G.~P. Fettweis, M.~C. Filippou, A.~Gati, M.-H. Hamon,
  M.~Hoffmann, M.~Latva-Aho, A.~Pärssinen, B.~Richerzhagen, H.~Schotten,
  T.~Svensson, G.~Wikström, H.~Wymeersch, V.~Ziegler, and Y.~Zou, ``6g vision,
  value, use cases and technologies from european 6g flagship project hexa-x,''
  \emph{IEEE Access}, vol.~9, pp. 160\,004--160\,020, 2021.

\bibitem{di2020smart}
M.~Di~Renzo, A.~Zappone, M.~Debbah, M.-S. Alouini, C.~Yuen, J.~De~Rosny, and
  S.~Tretyakov, ``Smart radio environments empowered by reconfigurable
  intelligent surfaces: How it works, state of research, and the road ahead,''
  \emph{IEEE Journal on Selected Areas in Communications}, vol.~38, no.~11, pp.
  2450--2525, 2020.

\bibitem{direnzo2021surveyRIS}
Y.~Liu, X.~Liu, X.~Mu, T.~Hou, J.~Xu, M.~Di~Renzo, and N.~Al-Dhahir,
  ``Reconfigurable intelligent surfaces: Principles and opportunities,''
  \emph{IEEE Communications Surveys Tutorials}, vol.~23, no.~3, pp. 1546--1577,
  2021.

\bibitem{10044963}
K.~Keykhosravi, B.~Denis, G.~C. Alexandropoulos, Z.~S. He, A.~Albanese,
  V.~Sciancalepore, and H.~Wymeersch, ``Leveraging ris-enabled smart signal
  propagation for solving infeasible localization problems: Scenarios, key
  research directions, and open challenges,'' \emph{IEEE Vehicular Technology
  Magazine}, vol.~18, no.~2, pp. 20--28, 2023.

\bibitem{8264743}
S.~Hu, F.~Rusek, and O.~Edfors, ``Beyond massive mimo: The potential of
  positioning with large intelligent surfaces,'' \emph{IEEE Transactions on
  Signal Processing}, vol.~66, no.~7, pp. 1761--1774, 2018.

\bibitem{9775078}
H.~Zhang, B.~Di, K.~Bian, Z.~Han, H.~V. Poor, and L.~Song, ``Toward ubiquitous
  sensing and localization with reconfigurable intelligent surfaces,''
  \emph{Proceedings of the IEEE}, vol. 110, no.~9, pp. 1401--1422, 2022.

\bibitem{Buzzi_RISforradar_letter}
S.~Buzzi, E.~Grossi, M.~Lops, and L.~Venturino, ``Radar target detection aided
  by reconfigurable intelligent surfaces,'' \emph{IEEE Signal Processing
  Letters}, vol.~28, pp. 1315--1319, 2021.

\bibitem{Buzzi_RISforradar_journal}
------, ``Foundations of mimo radar detection aided by reconfigurable
  intelligent surfaces,'' \emph{IEEE Transactions on Signal Processing},
  vol.~70, pp. 1749--1763, 2022.

\bibitem{Buzzi2022_activeRIS}
G.~Mylonopoulos, C.~D’Andrea, and S.~Buzzi, ``Active reconfigurable
  intelligent surfaces for user localization in mmwave mimo systems,'' in
  \emph{2022 IEEE 23rd International Workshop on Signal Processing Advances in
  Wireless Communication (SPAWC)}, 2022, pp. 1--5.

\bibitem{9508872}
A.~Elzanaty, A.~Guerra, F.~Guidi, and M.-S. Alouini, ``Reconfigurable
  intelligent surfaces for localization: Position and orientation error
  bounds,'' \emph{IEEE Transactions on Signal Processing}, vol.~69, pp.
  5386--5402, 2021.

\bibitem{9511765}
A.~Aubry, A.~De~Maio, and M.~Rosamilia, ``Ris-aided radar sensing in n-los
  environment,'' in \emph{2021 IEEE 8th International Workshop on Metrology for
  AeroSpace (MetroAeroSpace)}, 2021, pp. 277--282.

\bibitem{Zhang2022metalocalization}
H.~Zhang, H.~Zhang, B.~Di, K.~Bian, Z.~Han, and L.~Song, ``Metalocalization:
  Reconfigurable intelligent surface aided multi-user wireless indoor
  localization,'' \emph{IEEE Transactions on Wireless Communications}, vol.~20,
  no.~12, pp. 7743--7757, 2021.

\bibitem{Zhang2022metaradar}
------, ``Metaradar: Multi-target detection for reconfigurable intelligent
  surface aided radar systems,'' \emph{IEEE Transactions on Wireless
  Communications}, pp. 1--1, 2022.

\bibitem{Wang2022_location_awareness}
Z.~Wang, Z.~Liu, Y.~Shen, A.~Conti, and M.~Z. Win, ``Location awareness in
  beyond 5g networks via reconfigurable intelligent surfaces,'' \emph{IEEE
  Journal on Selected Areas in Communications}, vol.~40, no.~7, pp. 2011--2025,
  2022.

\bibitem{Dardari2023_twotimescale}
F.~Jiang, A.~Abrardo, K.~Keykhoshravi, H.~Wymeersch, D.~Dardari, and
  M.~Di~Renzo, ``Two-timescale transmission design and ris optimization for
  integrated localization and communications,'' \emph{IEEE Transactions on
  Wireless Communications}, pp. 1--1, 2023.

\bibitem{10096904}
P.~Zheng, H.~Chen, T.~Ballal, H.~Wymeersch, and T.~Y. Al-Naffouri,
  ``Misspecified cramér-rao bound of ris-aided localization under geometry
  mismatch,'' in \emph{ICASSP 2023 - 2023 IEEE International Conference on
  Acoustics, Speech and Signal Processing (ICASSP)}, 2023, pp. 1--5.

\bibitem{9838638}
Z.~Wang, Z.~Liu, Y.~Shen, A.~Conti, and M.~Z. Win, ``Source localization with
  intelligent surfaces,'' in \emph{ICC 2022 - IEEE International Conference on
  Communications}, 2022, pp. 895--900.

\bibitem{10146296}
G.~Oliveri, M.~Salucci, and A.~Massa, ``Generalized analysis and unified design
  of em skins,'' \emph{IEEE Transactions on Antennas and Propagation}, vol.~71,
  no.~8, pp. 6579--6592, 2023.

\bibitem{9625826}
D.~Dardari, N.~Decarli, A.~Guerra, and F.~Guidi, ``Los/nlos near-field
  localization with a large reconfigurable intelligent surface,'' \emph{IEEE
  Transactions on Wireless Communications}, vol.~21, no.~6, pp. 4282--4294,
  2022.

\bibitem{9650561}
M.~Luan, B.~Wang, Y.~Zhao, Z.~Feng, and F.~Hu, ``Phase design and near-field
  target localization for ris-assisted regional localization system,''
  \emph{IEEE Transactions on Vehicular Technology}, vol.~71, no.~2, pp.
  1766--1777, 2022.

\bibitem{9709801}
O.~Rinchi, A.~Elzanaty, and M.-S. Alouini, ``Compressive near-field
  localization for multipath ris-aided environments,'' \emph{IEEE
  Communications Letters}, vol.~26, no.~6, pp. 1268--1272, 2022.

\bibitem{9921216}
X.~Zhang and H.~Zhang, ``Hybrid reconfigurable intelligent surfaces-assisted
  near-field localization,'' \emph{IEEE Communications Letters}, vol.~27,
  no.~1, pp. 135--139, 2023.

\bibitem{Wymeersch2021}
K.~Keykhosravi, M.~F. Keskin, S.~Dwivedi, G.~Seco-Granados, and H.~Wymeersch,
  ``Semi-passive 3d positioning of multiple ris-enabled users,'' \emph{IEEE
  Transactions on Vehicular Technology}, vol.~70, no.~10, pp. 11\,073--11\,077,
  2021.

\bibitem{Wymeersch2022}
R.~Ghazalian, K.~Keykhosravi, H.~Chen, H.~Wymeersch, and R.~Jäntti,
  ``Bi-static sensing for near-field ris localization,'' in \emph{GLOBECOM 2022
  - 2022 IEEE Global Communications Conference}, 2022, pp. 6457--6462.

\bibitem{Zhang2023_targetRIS}
\BIBentryALTinterwordspacing
P.~Wang, W.~Mei, J.~Fang, and R.~Zhang, ``Target-mounted intelligent reflecting
  surface for joint location and orientation estimation,'' 2023. [Online].
  Available: \url{https://arxiv.org/abs/2301.09248}
\BIBentrySTDinterwordspacing

\bibitem{Wymeersch2022_wideband_RIS}
K.~Keykhosravi, M.~F. Keskin, G.~Seco-Granados, P.~Popovski, and H.~Wymeersch,
  ``Ris-enabled siso localization under user mobility and spatial-wideband
  effects,'' \emph{IEEE Journal of Selected Topics in Signal Processing},
  vol.~16, no.~5, pp. 1125--1140, 2022.

\bibitem{tagliaferri2023wideband}
D.~Tagliaferri, ``Wideband effects on near-field pose estimation of
  target-lodged ris,'' 2023.

\bibitem{Mizmizi2022_conformal}
M.~Mizmizi, R.~A. Ayoubi, D.~Tagliaferri, K.~Dong, G.~G. Gentili, and
  U.~Spagnolini, ``Conformal metasurfaces: a novel solution for vehicular
  communications,'' \emph{IEEE Transactions on Wireless Communications}, pp.
  1--1, 2022.

\bibitem{Tagliaferri2022_conformal6gnet}
D.~Tagliaferri, M.~Mizmizi, R.~A. Ayoubi, G.~G. Gentili, and U.~Spagnolini,
  ``Conformal intelligent reflecting surfaces for 6g v2v communications,'' in
  \emph{2022 1st International Conference on 6G Networking (6GNet)}, 2022, pp.
  1--8.

\bibitem{Mizmizi2022_conformal_GLOBECOM}
M.~Mizmizi, D.~Tagliaferri, M.~Khosronejad, L.~Resteghini, G.~G. Gentili,
  L.~Draghi, and U.~Spagnolini, ``Conformal metasurfaces for recovering dynamic
  blockage in vehicular systems,'' in \emph{GLOBECOM 2022 - 2022 IEEE Global
  Communications Conference}, 2022, pp. 6025--6030.

\bibitem{Wymeersch_CRB_radar_extendedvehicle}
N.~Garcia, A.~Fascista, A.~Coluccia, H.~Wymeersch, C.~Aydogdu, R.~Mendrzik, and
  G.~Seco-Granados, ``Cramér-rao bound analysis of radars for extended
  vehicular targets with known and unknown shape,'' \emph{IEEE Transactions on
  Signal Processing}, vol.~70, pp. 3280--3295, 2022.

\bibitem{9399297}
A.~F. Garc\'{i}a-Fern\'{a}ndez, J.~L. Williams, L.~Svensson, and Y.~Xia, ``A
  poisson multi-bernoulli mixture filter for coexisting point and extended
  targets,'' \emph{IEEE Transactions on Signal Processing}, vol.~69, pp.
  2600--2610, 2021.

\bibitem{naraghi1983geometric}
M.~Naraghi, W.~Stromberg, and M.~Daily, ``Geometric rectification of radar
  imagery using digital elevation models,'' \emph{Photogrammetric Engineering
  and Remote Sensing}, vol.~49, no.~2, pp. 195--199, 1983.

\bibitem{Desai2020TerahertzVA}
D.~Desai, I.~Gatley, C.~Bolton, L.~Rizzo, S.~Gatley, and J.~F. Federici,
  ``Terahertz van atta retroreflecting arrays,'' \emph{Journal of Infrared,
  Millimeter, and Terahertz Waves}, pp. 1--12, 2020.

\bibitem{9718037}
G.~Oliveri, F.~Zardi, P.~Rocca, M.~Salucci, and A.~Massa, ``Building a smart em
  environment - ai-enhanced aperiodic micro-scale design of passive em skins,''
  \emph{IEEE Transactions on Antennas and Propagation}, vol.~70, no.~10, pp.
  8757--8770, 2022.

\bibitem{9975205}
------, ``Constrained design of passive static em skins,'' \emph{IEEE
  Transactions on Antennas and Propagation}, vol.~71, no.~2, pp. 1528--1538,
  2023.

\bibitem{9580737}
G.~Oliveri, P.~Rocca, M.~Salucci, and A.~Massa, ``Holographic smart em skins
  for advanced beam power shaping in next generation wireless environments,''
  \emph{IEEE Journal on Multiscale and Multiphysics Computational Techniques},
  vol.~6, pp. 171--182, 2021.

\bibitem{14rel}
{3GPP TR 37.885 v15.3.}, ``Study on evaluation methodology of new {V2X} use
  cases for {LTE} and {NR} - {3rd Generation Partnership Project (3GPP)},''
  2019.

\bibitem{Folster2005}
F.~Folster and H.~Rohling, ``Data association and tracking for automotive radar
  networks,'' \emph{IEEE Transactions on Intelligent Transportation Systems},
  vol.~6, no.~4, pp. 370--377, 2005.

\bibitem{8730493}
K.~Granström, M.~Fatemi, and L.~Svensson, ``Poisson multi-bernoulli mixture
  conjugate prior for multiple extended target filtering,'' \emph{IEEE
  Transactions on Aerospace and Electronic Systems}, vol.~56, no.~1, pp.
  208--225, 2020.

\bibitem{5466116}
B.-N. Vo, B.-T. Vo, N.-T. Pham, and D.~Suter, ``Joint detection and estimation
  of multiple objects from image observations,'' \emph{IEEE Transactions on
  Signal Processing}, vol.~58, no.~10, pp. 5129--5141, 2010.

\bibitem{8443598}
B.~Wang, F.~Gao, S.~Jin, H.~Lin, G.~Y. Li, S.~Sun, and T.~S. Rappaport,
  ``Spatial-wideband effect in massive mimo with application in mmwave
  systems,'' \emph{IEEE Communications Magazine}, vol.~56, no.~12, pp.
  134--141, 2018.

\bibitem{ulaby2014microwave}
\BIBentryALTinterwordspacing
F.~Ulaby, D.~Long, C.~Elachi, and K.~Sarabandi, \emph{Microwave Radar and
  Radiometric Remote Sensing}.\hskip 1em plus 0.5em minus 0.4em\relax
  University of Michigan Press, 2014. [Online]. Available:
  \url{https://books.google.it/books?id=y6pZngEACAAJ}
\BIBentrySTDinterwordspacing

\bibitem{Bjornson2020}
O.~Ozdogan, E.~Bjornson, and E.~G. Larsson, ``Intelligent reflecting surfaces:
  Physics, propagation, and pathloss modeling,'' \emph{IEEE Wireless
  Communications Letters}, vol.~9, no.~5, pp. 581--585, 2020.

\bibitem{Ellingson2021}
S.~Ellingson, ``Path loss in reconfigurable intelligent surface-enabled
  channels,'' in \emph{2021 IEEE 32nd Annual International Symposium on
  Personal, Indoor and Mobile Radio Communications (PIMRC)}, 2021, pp.
  829--835.

\bibitem{TI_ref_MMWCAS}
T.~Instruments, ``Imaging radar using cascaded mmwave sensor reference
  design,'' available at: https://www.ti.com/lit/ug/tiduen5a/tiduen5a.pdf.

\bibitem{skolnik}
M.~Skolnik, \emph{Introduction to Radar Systems}.\hskip 1em plus 0.5em minus
  0.4em\relax London: McGraw-Hill Education, 2002.

\bibitem{wavefarer}
``{Remcom WaveFarer},'' \url{
  https://www.remcom.com/wavefarer-automotive-radar-software }, {Accessed on
  May 2023}.

\bibitem{wymeersch2023fundamental}
H.~Wymeersch, R.~Amiri, and G.~Seco-Granados, ``Fundamental performance bounds
  for carrier phase positioning in cellular networks,'' 2023.

\bibitem{BigS}
U.~{Spagnolini}, \emph{Statistical Signal Processing in Engineering}.\hskip 1em
  plus 0.5em minus 0.4em\relax John Wiley \& Sons Ltd, 2018.

\end{thebibliography}

\end{document}